\documentclass{aa}  
\usepackage{graphicx}
\usepackage{longtable}
\usepackage{tabularx}
\usepackage{txfonts}
\usepackage{hyperref}
\usepackage{scalerel}
\usepackage{lipsum}
\usepackage{subcaption}         
\usepackage{lscape}             
\usepackage{placeins}           
                                
\usepackage{xcolor}
\definecolor{dark-red}{rgb}{0.9,0.0,0.0}
\definecolor{dark-blue}{rgb}{0.15,0.15,0.9}
\definecolor{dark-green}{rgb}{0.15,0.8,0.15}
\definecolor{medium-blue}{rgb}{0,0,0.9}
\hypersetup{
    colorlinks, linkcolor=red,
    citecolor={dark-blue} , urlcolor={medium-blue}
}

\usepackage{comment}
\usepackage{amsmath, bm, nccmath}
\usepackage{subcaption}
\usepackage[export]{adjustbox}
\usepackage{float}
\usepackage{epstopdf}

\newcommand{\ceres}{\textsf{CERES}}

\newcommand{\mearth}{\ensuremath{{\rm M}_{\oplus}}}

\newcommand{\mstar}{\ensuremath{{\rm M}_{\star}}}

\newcommand{\msun}{\ensuremath{{\rm M}_{\odot}}}

\begin{document}

   \title{A transiting hot Jupiter with two outer siblings orbiting an intermediate-mass post main-sequence star}

\titlerunning{The TOI-375 planetary system}


\author{
Y. Reinarz\inst{1} \and
M. I. Jones\inst{2} \and
R. Brahm\inst{3,4} \and
N. Espinoza\inst{5} \and
M. Tala Pinto\inst{3} \and
T. Trifonov\inst{6,7} \and
A. Jordán\inst{3,4} \and
L. Acuña-Aguirre\inst{1} \and
T. Henning\inst{1} \and
F. Rojas\inst{8} \and
C. Ziegler\inst{10} \and
D. M. Conti\inst{14} \and
C. Brice\~{n}o\inst{9} \and
N. Law\inst{11} \and
A. W. Mann\inst{11} \and
K. A. Collins\inst{12} \and
J. M. Irwin\inst{13} \and
D. Charbonneau\inst{12}
}

\institute{
Max-Planck-Institut für Astronomie, Königstuhl 17, Heidelberg 69117, Germany\\
\email{reinarz@mpia.de} \and
European Southern Observatory, Alonso de Córdova 3107, Vitacura, Casilla 19001, Santiago, Chile \and
Facultad de Ingeniería y Ciencias, Universidad Adolfo Ibáñez, Av. Diagonal las Torres 2640, Peñalolén, Santiago, Chile \and
Millennium Institute for Astrophysics, Chile \and 
Space Telescope Science Institute, 3700 San Martin Drive, Baltimore, MD 21218, USA \and 
Landessternwarte, Zentrum f\"ur Astronomie der Universt\"at Heidelberg, K\"onigstuhl 12, 69117 Heidelberg, Germany \and
Department of Astronomy, Faculty of Physics, Sofia University ``St Kliment Ohridski'', 5 James Bourchier Blvd, 1164 Sofia, Bulgaria \and
Instituto de Astrofísica, Pontificia Universidad Católica de Chile, Av. Vicuña Mackenna 4860, Macul, Santiago, Chile \and
Cerro Tololo Inter-American Observatory, CTIO/AURA Inc., Chile \and
Department of Physics, Engineering and Astronomy, Stephen F. Austin State University, 1936 North St, Nacogdoches, TX 75962, USA \and
Department of Physics and Astronomy, The University of North Carolina at Chapel Hill, Chapel Hill, NC 27599-3255, USA \and
Center for Astrophysics \textbar \ Harvard \& Smithsonian, 60 Garden Street, Cambridge, MA 02138, USA \and
Institute of Astronomy, University of Cambridge, Madingley Road, Cambridge, CB3 0HA, UK \and 
American Association of Variable Star Observers, 185 Alewife Brook Parkway, Suite 410, Cambridge, MA 02138, USA
}

   \date{Received September 11, 2025}

\abstract
{
Exoplanetary systems with multiple giant planets present an opportunity to understand planet formation, migration processes, and long-term system-wide dynamical interactions. In particular, they provide constraints to distinguish between smooth disk-driven migration or more dynamically excited system evolution pathways.
We report the discovery and characterization of a unique multi-planet system hosting three gas giant planets, orbiting the post-main sequence star TOI-375. The innermost planet, TOI-375\,b, was initially detected by the \textit{TESS} mission, and then confirmed with photometric follow-up observations conducted using \textit{MEarth} and \textit{LCOGT}, and radial velocity measurements obtained with \textit{FEROS} and \textit{CHIRON}. The radial velocity data revealed the presence of two additional planetary candidates, TOI-375\,c and TOI-375\,d. We find that TOI-375\,b is a hot Jupiter with an orbital period of \ensuremath{9.45469 \pm 0.00002}\,days, mass \ensuremath{0.745 \pm 0.053\,M_\mathrm{J}}, radius \ensuremath{0.961 \pm 0.043\,R_\mathrm{J}}, and eccentricity \ensuremath{0.087 \pm 0.042}. The outer two planets, TOI-375\,c and TOI-375\,d, are warm-cold and cold Jupiters with orbital periods of \ensuremath{115.5^{+2.0}_{-1.6}}\,days and \ensuremath{297.9^{+28.9}_{-18.6}}\,days, and minimum masses of \ensuremath{2.11 \pm 0.22\,M_\mathrm{J}} and \ensuremath{1.40 \pm 0.28\,M_\mathrm{J}}, respectively. 
In terms of formation and overall system architecture, the physical properties of TOI-375\,b are consistent with the core accretion scenario, while the current configuration of the system could be explained by both disk-driven and high-eccentricity migration scenarios. The discovery of TOI-375 as the first known system hosting three or more fully evolved gas giants, with at least one transiting planet, makes it an excellent candidate for testing formation and migration theories.
}


\maketitle

\section{Introduction\label{sec:int}}

Hot Jupiters are highly irradiated gas giant exoplanets that orbit close to their host star with orbital periods less than 10 days. Due to strong observational biases favoring short-period, massive planets with large radius \citep{santerne2016}, they constitute a large fraction of the well characterized exoplanet sample despite their low occurrence rates
\citep{Stevenson1982,Fressin2013,mayor2011}.

Standard planet formation theories struggle to explain their existence at such close-in orbits. The two dominant formation models, core accretion \citep{Pollack1996} and gravitational instability \citep{Cameron1978}, both predict inefficient formation in the inner disk. In these regions, high temperatures inhibit the condensation of solids and prevent either the accumulation of massive cores or disk fragmentation \citep{Rafikov2005, Schlichting2014}.

The relatively high number of observed hot Jupiters, therefore, suggests that many did not form in situ, but rather underwent orbital migration. Two primary migration mechanisms are discussed in the literature. In disk-driven migration, torques between the forming planet and the protoplanetary gas disk lead to gradual inward movement \citep{lin1996}. Alternatively, high-eccentricity migration involves dynamical excitation of orbital eccentricity through mechanisms such as planet–planet scattering \citep{chatterjee2008} , Kozai–Lidov cycles induced by a stellar or planetary companion \citep{wu2003}, or secular chaos \citep{wu2011}, followed by tidal circularization close to the star. These scenarios imply different observable outcomes in terms of orbital eccentricity, obliquity, and system architecture. Disk-driven migration typically leads to low orbital eccentricities, as interactions with the protoplanetary disk dampen eccentricities \citep{goldreich1980,cresswell2008}. Planets that migrate through the disk tend to maintain low spin–orbit obliquities because the migration occurs within an aligned disk \citep{lin1996,dawson2018}, and the mutual inclinations between planets are expected to remain small \citep{kley2012}. These systems often host multiple, co-planar companions \citep{libert2011}, and show a smooth distribution of orbital periods as planets migrate inward \citep{mordasini:2009b}. The transit probability of outer planets remains high due to their low mutual inclinations \citep{DawsonChiang2014}, and minimal tidal heating or eccentricity evolution is expected over time \citep{kley2012}.

In contrast, high-eccentricity migration involves dynamical perturbations such as planet–planet scattering or Kozai–Lidov cycles that excite eccentric and inclined orbits before tidal forces circularize them \citep{chatterjee2008,naoz2016}. This process often produces high stellar obliquities, including retrograde orbits \citep{albrecht2012(banana)}. Systems that undergo this pathway frequently contain isolated hot Jupiters or have distant, inclined companions \citep{winn2010,bryan2016}. Their orbital period distribution shows a pile-up of hot Jupiters at short periods due to tidal circularization \citep{santerne2016}. The outer planets in such systems have a low transit probability because of their mutually inclined orbits \citep{read2017}, and the presence of tidal signatures or an eccentricity–age correlation may indicate ongoing orbital evolution \citep{dong2012}.

Well characterized multi-planetary systems provide key constraints to test these formation and migration pathways. The presence, proximity, and physical properties of planetary companions can help distinguish between smooth, disk-driven migration and more violent, dynamically excited evolution. As such, characterizing the architectures of systems hosting hot and cold Jupiters is crucial to understanding giant planet origins and orbital evolution.

The Transiting Exoplanet Survey Satellite \citep[\textit{TESS};][]{tess} is a space-based mission designed to search for exoplanets around bright, nearby stars by monitoring their light for periodic dips caused by planetary transits. By surveying nearly the entire sky, \textit{TESS} has enabled the identification of thousands of planetary candidates, many of which orbit stars suitable for follow-up observations. Its high-precision photometry and wide coverage make it particularly effective at detecting short-period planets and systems with multiple transiting planets, providing a valuable starting point for characterizing planetary architectures and informing studies of formation and migration.
In this work, we take a close look at TOI-375, an intermediate-mass (\mstar $\gtrsim$ 1.3 \msun) evolved star that hosts a  hot Jupiter detected by the \textit{TESS} mission and two outer gaseous companions, subsequently detected by radial velocity follow-up. 

The paper is structured as follows: In \S~\ref{sec:obs} we describe the observational material which gets used to perform a global modeling of the system as described in \S~\ref{sec:ana}. The results are then discussed in \S~\ref{sec:dis} and the conclusions are presented in \S~\ref{sec:con}.

\section{Observations} \label{sec:obs}

\subsection{\textit{TESS}}
\label{ssec:tess}
TOI-375 was observed by the Transiting Exoplanet Survey Satellite  over multiple sectors during its operational period. Initial observations were conducted at 30-minute cadence during Sectors 1 and 2, followed by 2-minute cadence observations in Sectors 13, 27, 28, 39, 66, 67, and 68. The \textit{TESS} data-validation reports revealed a transiting exoplanet candidate with a 9.45-day period and an approximate transit depth of 1000 ppm. The complete \textit{TESS} observation log is presented in Table~\ref{tab:tess_sectors}.

\begin{table}
\begin{tabular}{ccccc}
\hline
\hline
\textbf{Instrument} & \textbf{Start Date} & \textbf{End Date} & \textbf{Cad.} & $\mathbf{N^{observed}_{transits}}$ \\
\hline
\textit{TESS}-s1 & 25.07.2018 & 22.08.2018 & 30 min & 2\\ 
\textit{TESS}-s2 & 23.08.2018 & 18.09.2018 & 30 min & 2\\ 
\textit{TESS}-s13 & 04.06.2019 & 30.07.2019 & 2 min & 3\\ 
\textit{TESS}-s27 & 06.08.2019 & 04.09.2019 & 2 min & 3\\ 
\textit{TESS}-s28 & 05.09.2019 & 03.10.2019 & 2 min & 2\\ 
\textit{TESS}-s39 & 02.06.2020 & 01.07.2020 & 2 min & 3\\ 
\textit{TESS}-s66 & 02.06.2022 & 01.07.2022 & 2 min & 2\\ 
\textit{TESS}-s67 & 02.07.2022 & 01.08.2022 & 2 min & 2\\ 
\textit{TESS}-s68 & 02.08.2022 & 01.09.2022 & 2 min & 2\\ 
\hline
\end{tabular}
\caption{Observation details for \textit{TESS} sectors that contain observations of TOI-375.}
\label{tab:tess_sectors}
\end{table}


\subsection{Ground photometry}
\label{ssec:ground}

In addition to \textit{TESS} observations, ground-based photometric monitoring of TOI-375 was conducted using Las Cumbres Observatory Global Telescope \citep[\textit{LCOGT};][]{Brown:2013} 1.0 m network from 12.10.2019 to 26.10.2019, and \textit{MEarth-South} from 12.08.2019 to 21.08.2019. The \textit{LCOGT} 1\,m telescopes are equipped with $4096\times4096$ SINISTRO cameras having an image scale of $0\farcs389$ per pixel, resulting in a $26\arcmin\times26\arcmin$ field of view. The \textit{LCOGT} images were calibrated by the standard \textit{LCOGT} \textsf{BANZAI} pipeline \citep{McCully:2018} and differential photometric data were extracted using \textsf{AstroImageJ} \citep{Collins:2017}. While three transit ingresses and a single egress were observed according to the latest ephemeris, the shallow transit depth necessitates optimal observing conditions to extract meaningful information. Consequently, our analysis incorporates only two partial transit observations: an ingress detected by \textit{LCOGT} at the South African Astronomical Observatory (SAAO) on 09.09.2019, and another ingress observed by \textit{MEarth-South} on 21.08.2019. The complete ground-based observation log is presented in Table~\ref{tab:ground_observations}.

\begin{table}
\begin{tabular}{cccc}
\hline
\hline
\textbf{Night} &  \textbf{Facility} & \textbf{Cad.} &  \textbf{Obs.} \\
\hline
12.08.2019&$LCOGT_{CTIO}$ & 90s/45s & oot.\\
20.08.2019&$LCOGT_{SAAOz}$ & 90s &oot.\\
21.08.2019&$LCOGT_{CTIO}$ & 90s & first third, no oot.\\
\textbf{09.09.2019$^*$}&$LCOGT_{SAAOz}$& \textbf{90s} & \textbf{oot.} \textbf{ingress}\\
16.10.2019&$LCOGT_{SAAOz}$ &45s  & oot. ingress.\\
26.10.2019&$LCOGT_{SAAOz}$ &45s  & egress oot.\\
12.08.2019&$MEarth_{South}$   &120s &  oot. \\
\textbf{21.08.2019$^*$}&$MEarth_{South}$   &\textbf{120s} &  \textbf{oot.} \textbf{ingress}\\
\hline
\end{tabular}
\caption{Summary of ground-based observations. Here, oot denotes "out of transit". Observations at dates marked with an asterisk (*) were included in the final fit of the light curves.}
\label{tab:ground_observations}
\end{table}

\subsection{Spectroscopy}
We obtained 66 high-resolution spectra of TOI-375 using the \textit{FEROS} and \textit{CHIRON} spectrographs located in Chile between 2018 and 2019. A complete log of observations is provided in Table~\ref{tab:rvs}.

\subsubsection{\textit{FEROS}}
The majority of our spectroscopic dataset consists of observations conducted with the \textit{FEROS} spectrograph \citep{kaufer:99}, mounted on the MPG 2.2m telescope at La Silla Observatory. 
Between  07.02.2019 and 17.11.2019, we obtained 49 observations using simultaneous calibration mode, where a thorium-argon (ThAr) lamp was observed through the secondary fiber to track instrumental drifts caused by environmental variations during science exposures. 

The data were reduced using the \ceres\ pipeline \citep{brahm:2017:ceres}, which delivers radial velocities corrected for both instrumental drift and Earth's motion. These velocities were derived using the cross-correlation technique, employing a G2-type binary mask as the template spectrum. The pipeline additionally computes bisector span measurements from the cross-correlation peak and provides preliminary stellar parameters through comparison of the continuum-normalized spectrum with a grid of synthetic spectra. The latest version of \ceres\ also includes stellar activity indicators.

\subsubsection{\textit{CHIRON}}
We also collected 16 spectra using the \textit{CHIRON} high-resolution spectrograph \citep{tokovinin:2013} between 2018-10-21 and 2019-10-15. \textit{CHIRON} is mounted on the SMARTS 1.5\,m telescope at the Cerro Tololo Inter-American Observatory and is fed by an octagonal multi-mode optical fiber. We utilized the image slicer mode, which provides high throughput at a spectral resolution of R $\approx$ 80,000. Our exposure times ranged from 750 to 1200\,s, yielding signal-to-noise ratios (SNR) per extracted pixel between $\approx$ 20\,-\,35. 

The radial velocities were computed via cross-correlation between individual spectra and a high-resolution stellar template created by co-adding all observations of the target. In contrast to \textit{FEROS}, \textit{CHIRON} lacks simultaneous calibration capability. Therefore, we obtained Th-Ar lamp spectra before science observations to correct for instrumental drift. This methodology achieves long-term RV stability of $<$ 10\,m\,s$^{-1}$ for bright targets (t$_{\rm exp} <$ 60\,s) and $<$ 15\,m\,s$^{-1}$ for fainter objects (t$_{\rm exp} <$ 1800\,s). For detailed methodology, see \citet{wang:2019} and \citet{jones:2019}. A subset of observations obtained during August-September 2019 were acquired at lower spectral resolution and therefore required independent reduction procedures. The complete radial velocity measurements are presented in Table~\ref{tab:rvs}.

\section{Analysis} 
\label{sec:ana}

\subsection{Stellar parameters}
\label{ssec:stellarpars}

We used \textsf{Species} \citep{soto:2018} to derive the stellar atmospheric and bulk parameters, using a high signal-to-noise-ratio template, produced by co-adding the individual \textit{FEROS} data. This code has been adapted to robustly derive the stellar parameters of evolved stars \citep{Soto2021}. 
Briefly, \textsf{Species} measures the equivalent widths of the iron line and solves the radiative transfer equation under the assumption of local thermodynamic equilibrium to obtain the effective temperature ($T_{\mathrm{eff}}$), surface gravity ($\log g$), metallicity ([Fe/H]) and microturbulent velocity ($\xi_t$). From these values, \textsf{Species} estimates the chemical abundances, projected rotation velocity ($v \sin i$), and macroturbulent velocity, and, through Bayesian interpolation on a grid of MIST isochrones \citep{dotter:2016}, the stellar mass, radius, and age. For this work, we modified the code to incorporate parallaxes, proper motions, and magnitudes from \textit{Gaia}~DR3, and we increased the default priors range to allow for higher extinction values (A$_V$). This was needed given the large interstellar absorption in the line-of-sight. The resulting stellar properties are displayed in Table \ref{tab:stprops}.

\begin{figure}
    \centering
    \includegraphics[width=1.1\linewidth]{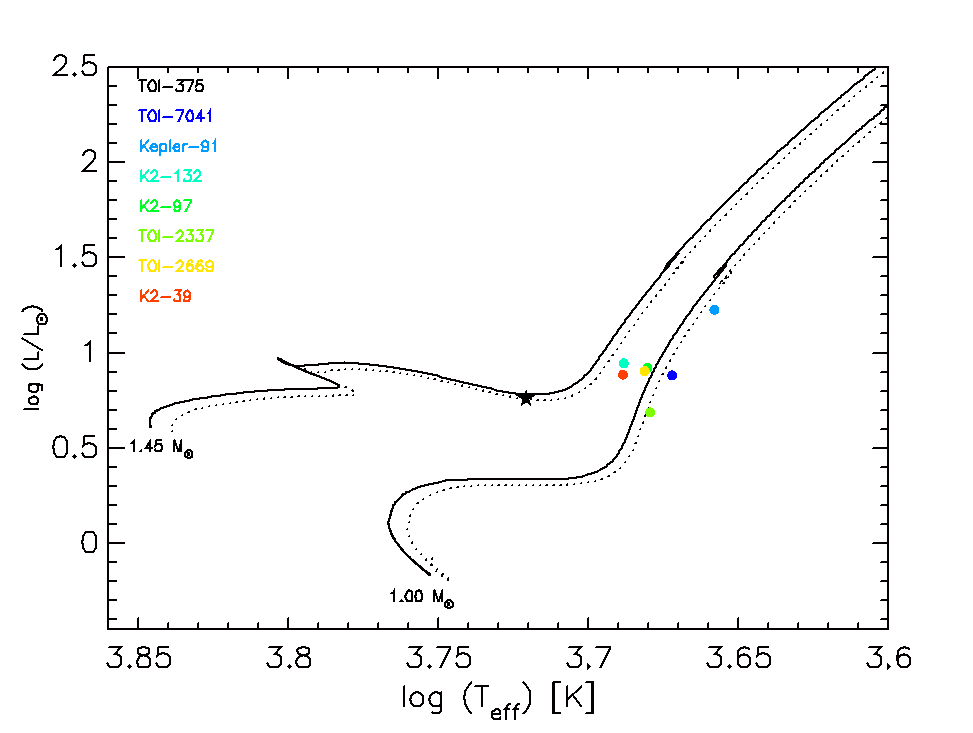}
    \caption{Position of TOI-375 in the HR diagram (black star), using the stellar parameters derived here (see Table 3). For comparison, the position of known giant stars hosting transiting giant planets are overplotted. Two different \textsf{PARSEC} models \citep{parsec} for 1.0 and 1.45 \msun, are overplotted. The solid and dotted lines correspond to Z$_\star$=0.017 and 0.020, respectively.}
    \label{fig:HR_diagram}
\end{figure}

Figure \ref{fig:HR_diagram} shows the position of TOI-375 in the HR diagram, along with different PARSEC evolutionary models, being the 1.45 \msun\, model with Z = 0.020 (corresponding to [Fe/H] = +0.11 dex), the closest one to its position in the grid. As can be seen, this star is currently at the end of the sub-giant branch phase, hence rapidly shrinking the H-burning shell around the (nearly) iso-thermal He-rich core, and about to start climbing the red giant branch.

\begin{table}
\caption{Stellar properties of TOI-375}
\label{tab:stprops}
\centering
\begin{tabular}{lcc}
\hline\hline
Parameter & Value & Reference \\
\hline
Names & TIC 280097543 & \textit{TESS} \\
      & 2MASS J03083602-7722598 & 2MASS \\
      & TYC 9366-00188-1 & TYCHO \\
      & WISE J030835.98-772259.8 & WISE \\
RA (J2000) & 03h08m35.98s & \\
DEC (J2000) & -77d22m59.98s & \\
$\mu_\alpha$ (mas yr$^{-1}$) & $-4.6136 \pm 0.0125$ & Gaia DR3 \\
$\mu_\delta$ (mas yr$^{-1}$) & $-4.2989 \pm 0.0164$ & Gaia DR3 \\
$\pi$ (mas) & $2.4779 \pm 0.0123$ & Gaia DR3 \\
\hline
\textit{TESS} (mag) & $10.4471 \pm 0.006$ & \textit{TESS} \\
$G$ (mag) & $11.0470 \pm 0.0002$ & Gaia DR3 \\
$B_P$ (mag) & $11.5421 \pm 0.0006$ & Gaia DR3 \\
$R_P$ (mag) & $10.3895 \pm 0.0006$ & Gaia DR3 \\
$J$ (mag) & $9.633 \pm 0.026$ & 2MASS \\
$H$ (mag) & $9.172 \pm 0.025$ & 2MASS \\
$K_s$ (mag) & $9.084 \pm 0.023$ & 2MASS \\
\hline
$L_\star$ ($L_\odot$) & $5.754^{+0.282}_{-0.283}$ & This work \\
$T_{\mathrm{eff}}$ (K) & $5259.9^{+136.9}_{-135.3}$ & This work \\
$\log g$ (dex) & $3.672 \pm 0.011$ & This work \\
$\mathrm{[Fe/H]}$ (dex) & $0.084^{+0.047}_{-0.045}$ & This work \\
$v \sin i$ (km s$^{-1}$) & $2.97 \pm 0.30$ & This work \\
$M_\star$ ($M_\odot$) & $1.441^{+0.076}_{-0.075}$ & This work \\
$R_\star$ ($R_\odot$) & $2.900^{+0.129}_{-0.127}$ & This work \\
Age (Gyr) & $2.94^{+0.60}_{-0.60}$ & This work \\
$\rho_\star$ (g cm$^{-3}$) & $0.083 \pm 0.012$ & This work \\
\hline
\end{tabular}
\end{table}

\subsection{Nearby companions}
\begin{figure}
    \centering
    \includegraphics[width=\linewidth]{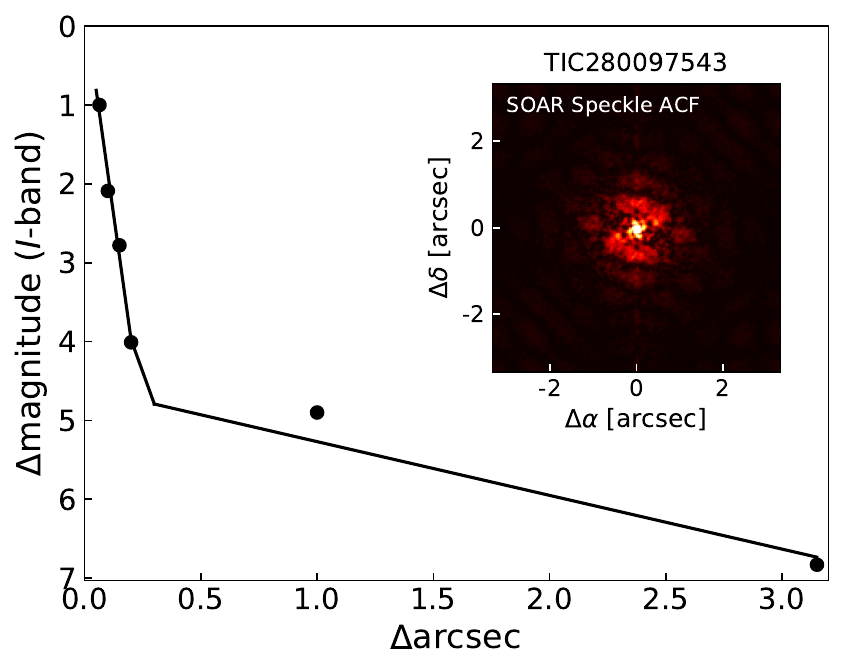}
    \caption{Speckle autocorrelation function for TOI-375, obtained with the \textit{SOAR} telescope. The black dots represent the $5\sigma$ contrast curve, and the solid line shows a linear fit to the data at separations smaller and larger than $\sim 0.2$ arcseconds.}
    \label{fig:soar}
\end{figure}

Since the planetary radius is determined from the relative stellar flux decrease during transit, accounting for contamination from nearby stars is crucial. Without this correction, the planetary radius can be underestimated due to overestimating the host star's contribution to the out-of-transit baseline flux. This contamination effect, known as dilution, can be accounted for by incorporating a dilution factor in the light curve model, which represents the fraction of the out-of-transit flux originating from the host star.\\ 
To assess possible contamination, we examined Target Pixel File (TPF) images with \textsf{tpfplotter}, utilizing stellar positions and magnitudes from the \textit{Gaia} catalog (Fig. \ref{fig:tpf}).  We also searched for closer stellar companions to TOI-375 using speckle imaging with the 4.1\,m Southern Astrophysical Research (\textit{SOAR}) telescope \citep{Tokovinin_2018}. The target was observed with HRCam on 2019 February 2. This observation was sensitive to companions up to 4.9 magnitudes at 1 arcsecond \citep{Ziegler_2019,Ziegler_2021}.  The $5\sigma$ detection limits and the speckle autocorrelation function from this observation are shown in Figure~\ref{fig:soar}. No nearby sources were detected within 3 arcseconds of TOI-375.\\

Both analyses show no evidence of contamination within the photometric aperture in any \textit{TESS} sector, allowing us to adopt a dilution factor of unity in subsequent analysis.

\subsection{Light curve analysis}
We analyzed photometric data from both space-based \textit{TESS} observations and ground-based facilities. For \textit{TESS} Sectors 1 and 2, we extracted light curves from the full frame images using \textsf{tesseract}, while for Sectors 13 through 68, we utilized the Pre-search Data Conditioning Simple Aperture Photometry (PDC-SAP) flux retrieved from the MAST portal. The photmoetric data from \textit{LCOGT} was automatically reduced using the BANZAI pipeline \citep{banzai}, which outputs the light curve for different aperture sizes. Since no contamination from nearby sources was present, we selected the largest available aperture of 6 arcseconds. The \textit{MEarth-South} observations were processed following the reduction techniques described in \citet{2007MNRAS.375.1449I} and \citet{2001NewAR..45..105I}. The dataset includes simultaneous aperture photometry from telescopes 11 through 17, which we combined into a single light curve for our analysis.

To search for periodic transit signals, we first detrended the \textit{TESS} light curves using a linear model and performed a Transit Least Squares (TLS) \citep{Hippke_2019} analysis using \textsf{exostriker} \citep{2019ascl.soft06004T}. This analysis revealed a strong signal with a Signal Detection Efficiency (SDE) of approximately 73, corresponding to the 9.45-day period planet previously reported in the \textit{TESS} data validation reports. While additional peaks were detected in the periodogram, we identified these as harmonics of the primary signal.

After obtaining a preliminary transit model for the 9.45-day planet candidate, we conducted a subsequent TLS analysis on the residuals using \textsf{exostriker} to search for additional periodic signals. This analysis revealed no additional signals that could indicate the presence of other transiting planets. However, we identified a potential single transit event in Sector 2 at around BJD 2458365.82 (Fig. \ref{fig:transit_models}). This region of the light curve exhibits particularly high noise levels, and requires further observations to confirm whether this signal corresponds to a planetary transit. As such, we do not include it in our final analysis.

\subsection{Radial-velocity analysis}
\begin{figure}
\includegraphics[width=\columnwidth]{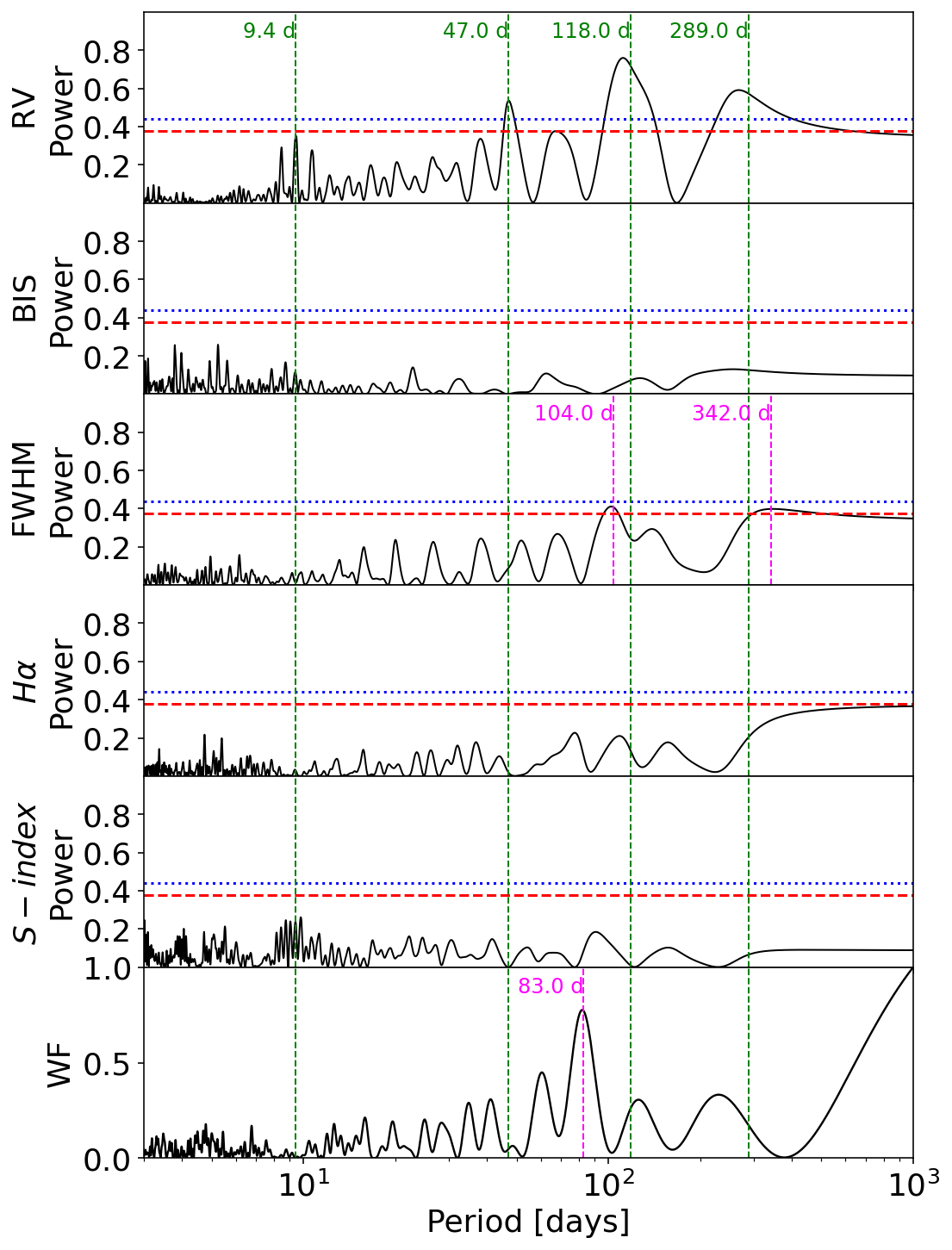}
\caption{Periodogram of the radial velocities (top) and activity indices (middle), along with the window function of the observations (bottom) for the \textit{FEROS} data of TOI-375. False-alarm probabilities are indicated by horizontal lines: dashed red for 1\% FAP and dotted blue for 0.1\% FAP on each panel. Green vertical lines mark peaks in the radial-velocity periodogram with FAP below 1\% or corresponding to a confirmed a priori planetary signal, such as the 9.4-day signal. In contrast, magenta vertical lines indicate signals that are unlikely to be of planetary origin.}
\label{fig:periodogram-toi375}
\end{figure}

The Generalized Lomb–Scargle (GLS) periodograms \citep{GLS} of the \textit{FEROS} radial velocity data reveal strong periodic signals at 289, 118, and 47 days, all well below the 0.1\% false-alarm probability (FAP) (Fig.~\ref{fig:periodogram-toi375}). An additional weak signal is detected at 9.43 days, corresponding to the transiting planet identified by \textit{TESS}. In contrast, standard stellar activity indicators, including the bisector (BIS) \citep{Queloz2001,Santos_2010}, H$\alpha$ \citep{Robertson_2013}, and chromospheric indices such as the S-index \citep{Noyes_1984}, show no significant power at these periods, suggesting that the radial velocity variations are unlikely to be driven by stellar activity. The FWHM exhibits peaks between the 1\%-0.1\% FAP range at around 104 and 342 days, but these signals are much weaker than those observed in the radial velocities at similar periods. The window function (WF) shows a peak at 83 days, which does not coincide with any of the radial velocity signals, indicating that the observed RV variations are unlikely to arise from the temporal sampling of the data.

To fit the radial velocities, we use the \textsf{juliet} library \citep{espinoza:juliet}. To select the best-fitting model, we employed a Bayesian model comparison approach. We computed the Akaike Information Criterion (AIC) and the Bayesian Information Criterion (BIC) for 16 different models that combine circular and eccentric orbits for planet~c (with orbital periods between 50 and 150~days) and planet~d (between 200 and 450~days). We also tested for the presence of a linear trend in the data. The best-fitting model includes no linear trend and adopts circular orbits for the outer planets, suggesting that the available data are insufficient to constrain their eccentricities. Additionally, we tested the presence of a planet corresponding to the 47-day signal, but this peak in the GLS periodogram corresponds to a harmonic of the 9.45-day planet and does not indicate a real planetary signal.

\subsection{Global modeling}
We used  \citep{espinoza:juliet} to jointly model the radial velocities and light curves of TOI-375. We adopted non-informative priors for all parameters except for the stellar density, for which we used a normal prior centered on the value derived with \textsf{Species}, with a standard deviation corresponding to its reported uncertainty. The full list of priors is presented in table \ref{table:full}.\\
In order to remove systematic effects from the light curves, we implemented different de-trending techniques for each instrument simultaneously during the global modeling of the system.
For \textit{TESS} Sectors 1 and 2, we implemented Gaussian processes using a Celerite Matérn kernel \citep{exoplanet:foremanmackey18} to account for correlated noise in the data. For the remaining \textit{TESS} sectors (13-68), we utilized the PDC-SAP flux. These light curves have already undergone systematic correction through the PDC pipeline and we only normalized it by the median of the baseline flux. The \textit{LCOGT} observations from SAAO were detrended using a two-dimensional linear model with time and airmass as regressors, accounting for both temporal trends and atmospheric effects. For \textit{MEarth} data, we implemented a two-dimensional linear model using time and FWHM as regressors, supplemented by Gaussian processes with a Matérn kernel using time as the regressor to capture additional correlated noise patterns.\\

\begin{figure}[]
    \centering
    \includegraphics[width=\columnwidth]{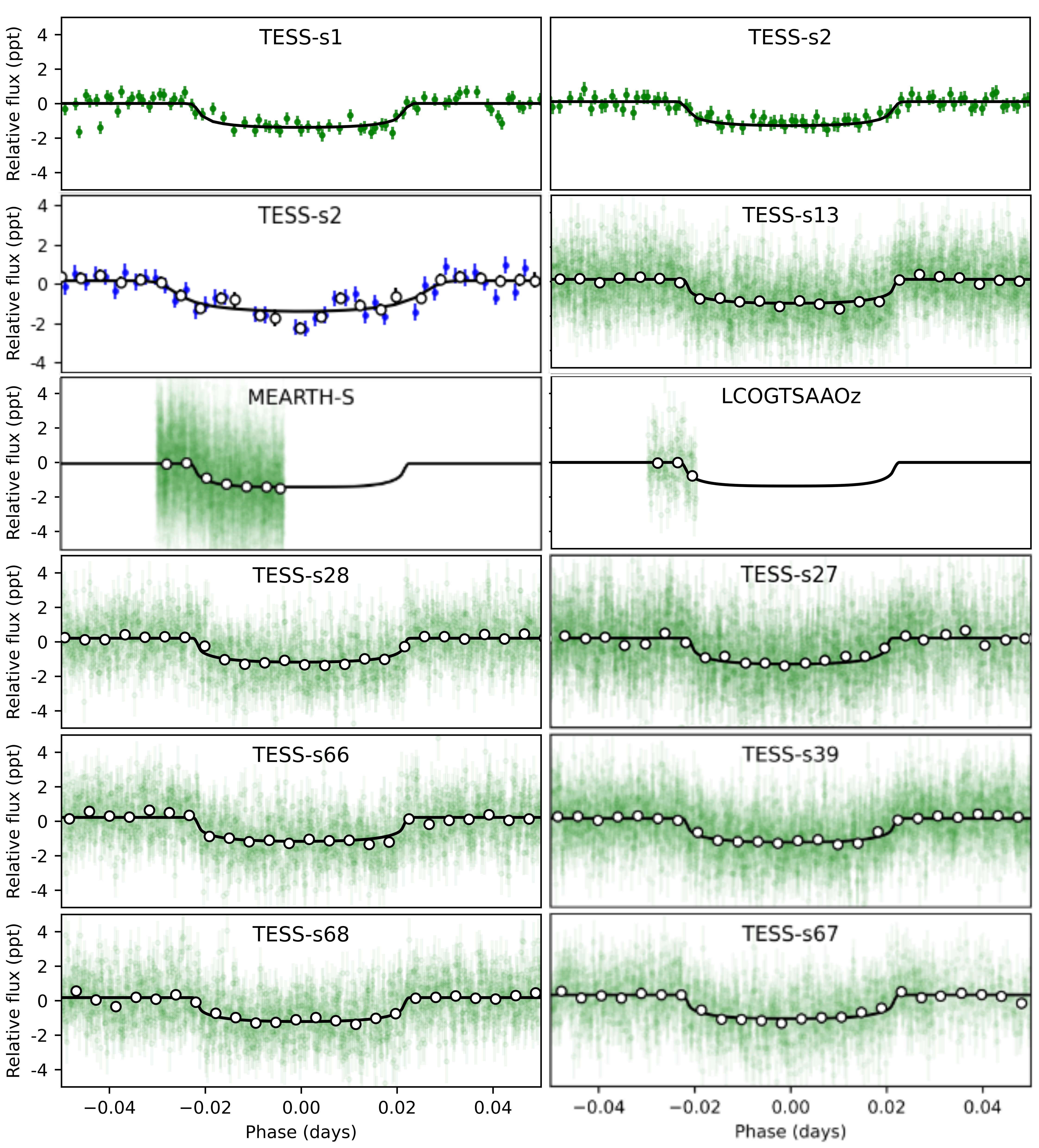}
    \caption{In green, we show the Phase-folded light curves from each instrument (or sector in the case of \textit{TESS}) with the best-fit transit model for TOI-375 b derived using \textsf{juliet}. In blue, we show the potential single transit event of one of the outer planets on \textit{TESS} sector 2. Data points are shown in their original time sampling (small green or blue points) and binned in phase with 6-minute intervals (white markers). The solid black line represents the best-fit transit model. We do not include the transit of TOI-375 c in the final model as there is not enough evidence to support it is from true planetary origin.}
    \label{fig:transit_models}
\end{figure}

\begin{figure*}[]
\includegraphics[trim={3cm 1cm 3cm 2cm},clip,width=1\textwidth]{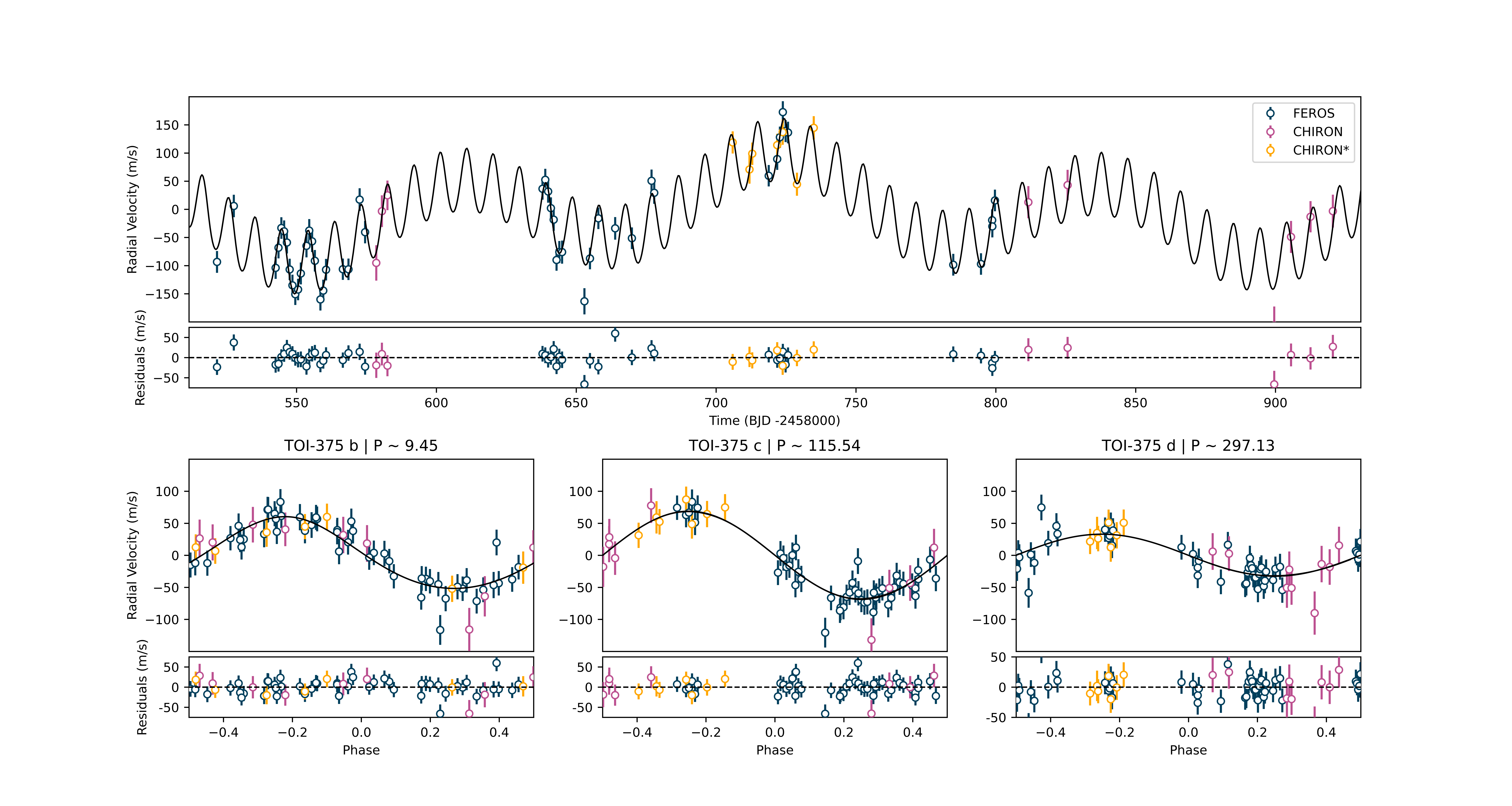}
\caption{(\textit{Top}) Radial velocities along with the median model (black line) for TOI-375. Our model is composed of 3 different Keplerian components. (\textit{Bottom}) Keplerian components of the full model presented in the top panel. The 9.4-day variation is from TOI-375 b, the transiting exoplanet detected by \textit{TESS}. The two additional 118.7 and 289.9-day signals observed in our radial-velocities have been identified as likely candidate exoplanets (c and d) given no obvious stellar activity origin.
}
\label{fig:rvs375}
\end{figure*}

To select the limb darkening law we adopted the logarithmic limb darkening law based on the recommendations listed in \citet{Espinoza_2016}, who studied the performance of different laws across a range of stellar and planetary parameters. For stars with effective temperatures around $5100\ \mathrm{K}$, planet-to-star radius ratios near $0.035$, and impact parameters between $0.35$ and $0.55$, the logarithmic law was found to perform optimally. Since limb darkening is wavelength/filter dependent, we provide independent priors for \textit{LCOGT}, \textit{MEarth}, and \textit{TESS}, but group all \textit{TESS} sectors together as they should exhibit consistent limb darkening behavior.\\
The posteriors of the physical parameters and ephemeris for the TOI-375 a, b and c are presented in Table \ref{table:results}. The  best-fit lines for the transit and radial velocity parts of the model are shown in Figures \ref{fig:transit_models} and \ref{fig:rvs375}, respectively.

\label{sec:glob}

\section{Discussion} \label{sec:dis}

\begin{table}[]
\centering
\begin{tabular}{l@{\hspace{70pt}}lr}
\hline
\hline
TOI-375{}b&$P(days)$&$9.45469^{+0.00002}_{-0.00002}$\\
&$t_0(BJD*)$&$329.291^{+0.002}_{-0.002}$\\
&$p(R_*)$&$0.0340^{+0.0004}_{-0.0004}$\\
&$b$&$0.1767^{+0.1387}_{-0.1137}$\\
&$K(ms^{-1})$&$56.27^{+3.37}_{-3.40}$\\
&$e$&$0.087^{+0.041}_{-0.042}$\\
&$\omega (deg)$&$9.107^{+25.105}_{-27.852}$\\
&$a(au)$&$0.099^{+0.002}_{-0.002}$\\
&$R_p(R_{J})$&$0.961^{+0.044}_{-0.043}$\\
&$M_p(M_J)$&$0.745^{+0.052}_{-0.053}$\\
&$T_{eq}(K)$&$1373^{+47}_{-45}$ \\
\hline
TOI-375{}c&$P(days)$&$115.5^{+2.0}_{-1.6}$\\
&$t_0(BJD*)$&$520.568^{+2.103}_{-2.242}$\\
&$K(ms^{-1})$&$69.18^{+6.29}_{-6.48}$\\
&$a(au)$&$0.524^{+0.011}_{-0.011}$\\
&$M_psin(i)(M_J)$& $2.106^{+0.223}_{-0.212}$\\
&$T_{eq}(K)$&$596^{+21}_{-21}$\\
\hline
TOI-375{}d&$P(days)$&$297.9^{+28.9}_{-18.6}$\\
&$t_0(BJD*)$&$790.670^{+10.790}_{-10.264}$\\
&$K(ms^{-1})$&$33.34^{+7.27}_{-6.51}$\\
&$a(au)$&$0.984^{+0.062}_{-0.044}$\\
&$M_psin(i)(M_J)$&$1.402^{+0.311}_{-0.276}$\\
&$T_{eq}(K)$&$434^{+19}_{-19}$\\
\hline
\end{tabular}
\caption{Physical parameters and ephemeris for TOI-375 b,c and d. $BJD*=BJD-2458000$}.
\label{table:results}
\end{table}

\begin{figure}[h]
    \centering
    \includegraphics[width=1\linewidth]{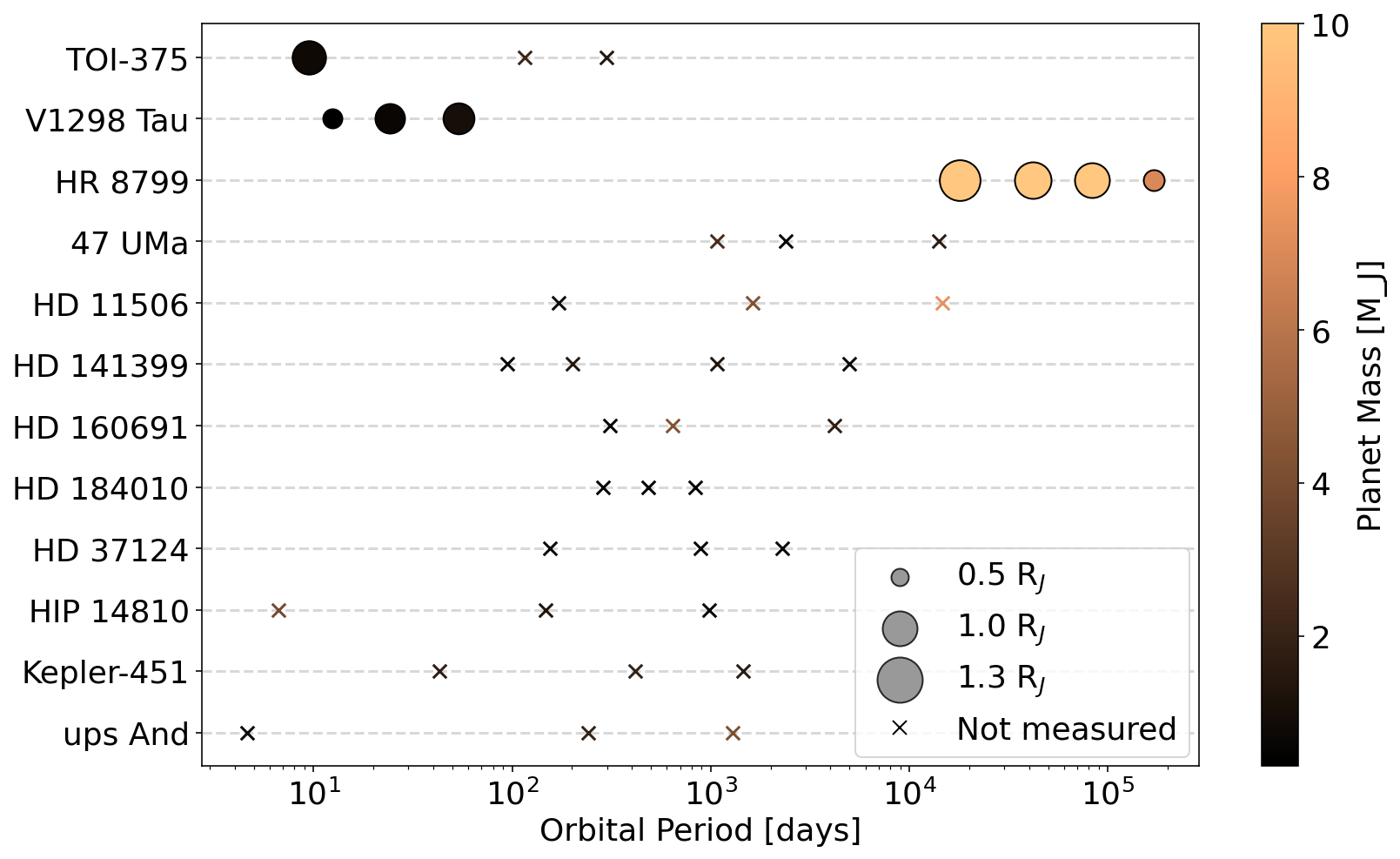}
    \caption{Systems with three or more planets in the mass range $0.3,M_J < M < 30,M_J$. Planets shown with a cross lack radius measurements and the mass used corresponds to the minimum mass ($Msin(i)$). TOI-375 represents the second ever discovery of a multi-jupiter-mass system where at least one planet’s transit has been detected.}
    \label{fig:muli_jupiter_systems}
\end{figure}

\subsection{System dynamical history}

Systems hosting three or more Jupiter-mass planets are uncommon, and cases where at least one of them transits are even rarer. To date, only one other such system, V1298 Tau \citep{David_2019,Feinstein_2022}, has been reported (Fig.~\ref{fig:muli_jupiter_systems}). We note, however, that V1298 Tau is very young ($\sim$20-30 Myr), and the gas giants in this system are more likely super-Earths or mini-Neptunes still in the process of contraction, mass loss, or both \citep{Thao_2024,Barat_2025}. This makes TOI-375 especially valuable for testing theories of system evolution and migration.

The low eccentricity of TOI-375b is consistent with disk-driven migration, where the planet forms beyond the snow line and migrates inward through the protoplanetary disk while maintaining low eccentricity and inclination \citep{lin1996,kley2012}. If the outer planets also have low eccentricities, which is a reasonable assumption given the current data, the system architecture would further support this formation pathway. High-eccentricity migration mechanisms, such as planet–planet scattering or Kozai–Lidov cycles, are expected to produce systems with higher eccentricities, misaligned orbits, and fewer nearby companions \citep{chatterjee2008,naoz2016}.

The lack of evidence for strong radial velocity trends or orbital instability suggests that the system is dynamically stable. However, without precise measurements of eccentricity and mutual inclinations, it is not possible to rule out past dynamical interactions such as mild scattering or secular perturbations.

Overall, the data are consistent with a formation scenario involving disk-driven migration for TOI-375b, accompanied by two outer giant planets on wider orbits. Additional observations, especially those that can constrain orbital inclinations and spin–orbit alignment, would help clarify the system’s dynamical history.

\subsection{TOI-375 b interior modeling}

\begin{figure}
    \centering
    \includegraphics[width=1\linewidth]{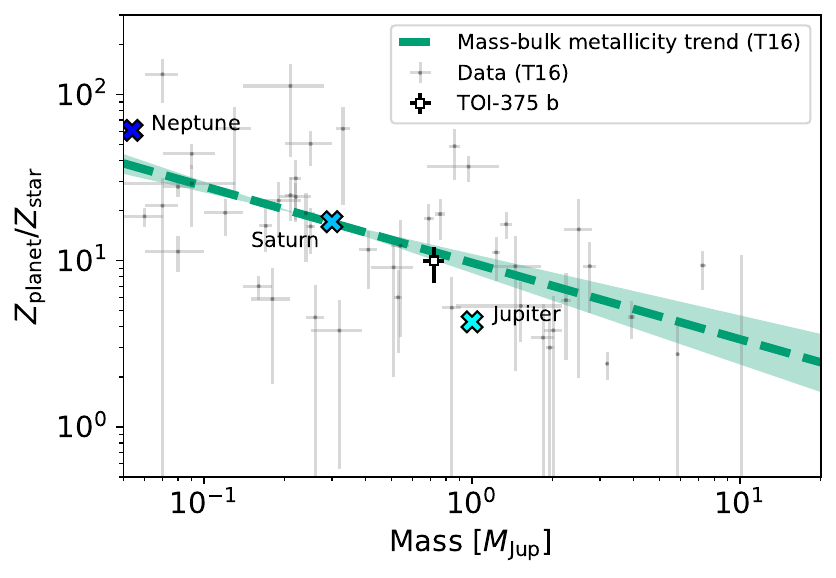}
    \caption{Metal enrichment as a function of mass for gas giants. The mass-metallicity trend and sample data are provided by \cite[T16][]{Thorngren_2016}}
    \label{fig:mass-met}
\end{figure}

\begin{figure}
    \centering
    \includegraphics[width=0.65\linewidth,angle=90]{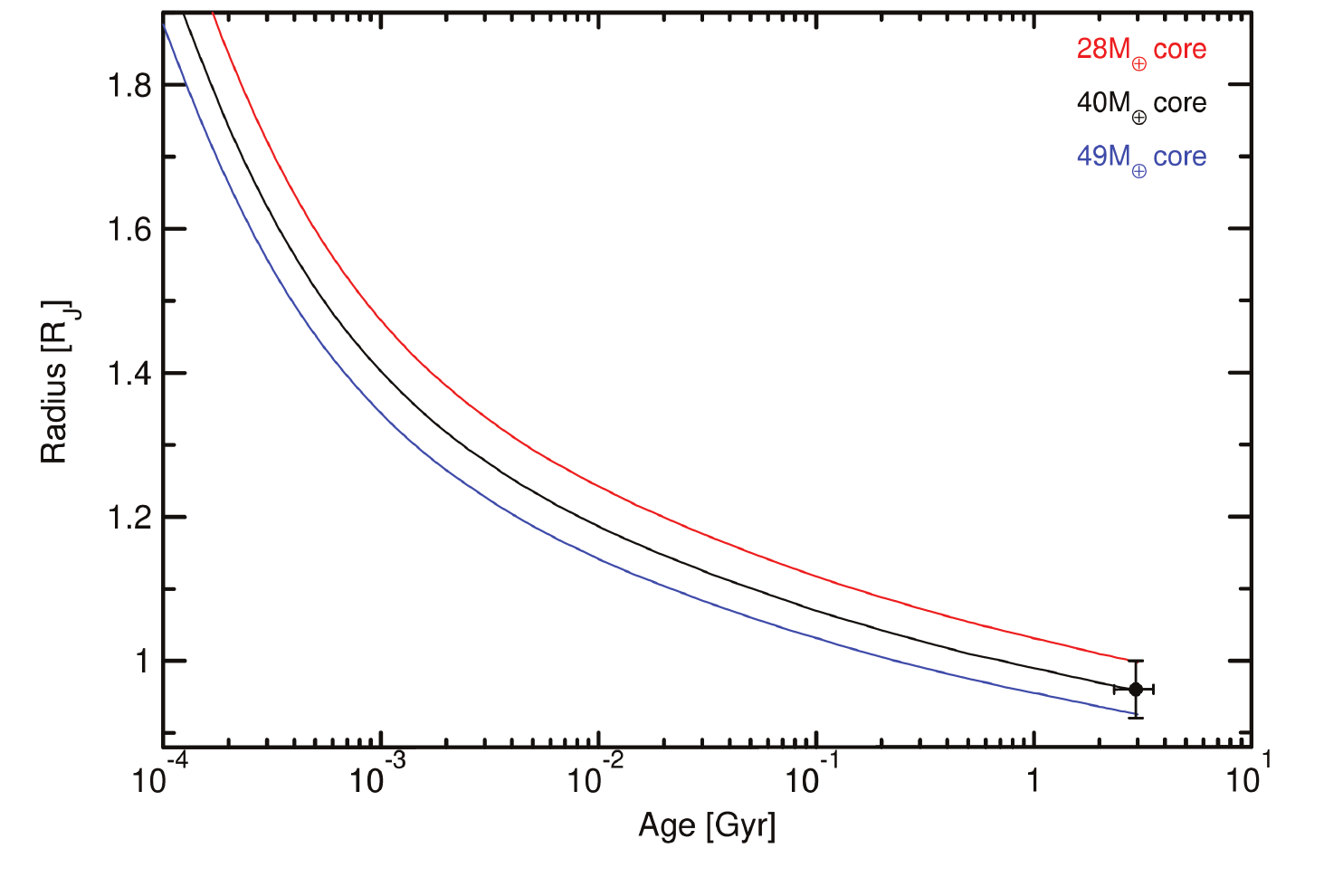}
    \caption{Position of TOI-375\,$b$ in the age-radius diagram (black dot) and planet evolutionary models with core masses of 28, 49, and 49\,\mearth. }
    \label{fig:mesa}
\end{figure}

We computed interior models of TOI-375 b using the Modules for Experiments in Stellar Astrophysics (\textsf{MESA}; \citealt{Paxton_2011}), closely following the methodology described in 
\citet{Jones_2024} and \citet{Tala2025}. The interior of the planet is modeled with a rocky-core surrounded by a gas rich H/He envelope. For this, we assumed an inert homogeneous core composed of a 1:1 mixture of rock and ice, whose density is computed using the \citet{Hubbard1989} $\rho-P$ tables (in this case $\rho_c$ = 9 ${\rm g\,cm^{-3}}$). The metallicity of the envelope is assumed to be the same as the host star ($Z_{\mathrm{env}} = 0.015$). Finally, the instant irradiation level received by the planet is also updated in steps of 250 Myr, using the PARSEC model that better matches the host star parameters. 
Figure~\ref{fig:mesa}, shows the position of TOI-375 b in the age-radius diagram. Different models with core masses between 28\,-\,49 \mearth\, that reproduce its position within 1$\sigma$ are overplotted. This corresponds to a planet metallicity of $Z_{\rm p}=0.18^{+0.04}_{-0.05}$, hence a planet heavy-element enrichment of $Z_{\rm p}/Z_\star$ = 10.0$^{+2.0}_{-2.7}$, in agreement with the core-accretion model of planet formation (Fig. \ref{fig:mass-met}). Also, despite the high irradiation level currently received by the planet of f$_p \sim$ 8$\times 10^8$ [erg\,s$^{-1}$\,cm$^{-2}$], no evidence of (re-)inflation is observed \citep{demory:2011,grunblatt16}.   

\begin{figure}
    \centering
    \includegraphics[width=0.65\linewidth,angle=90]{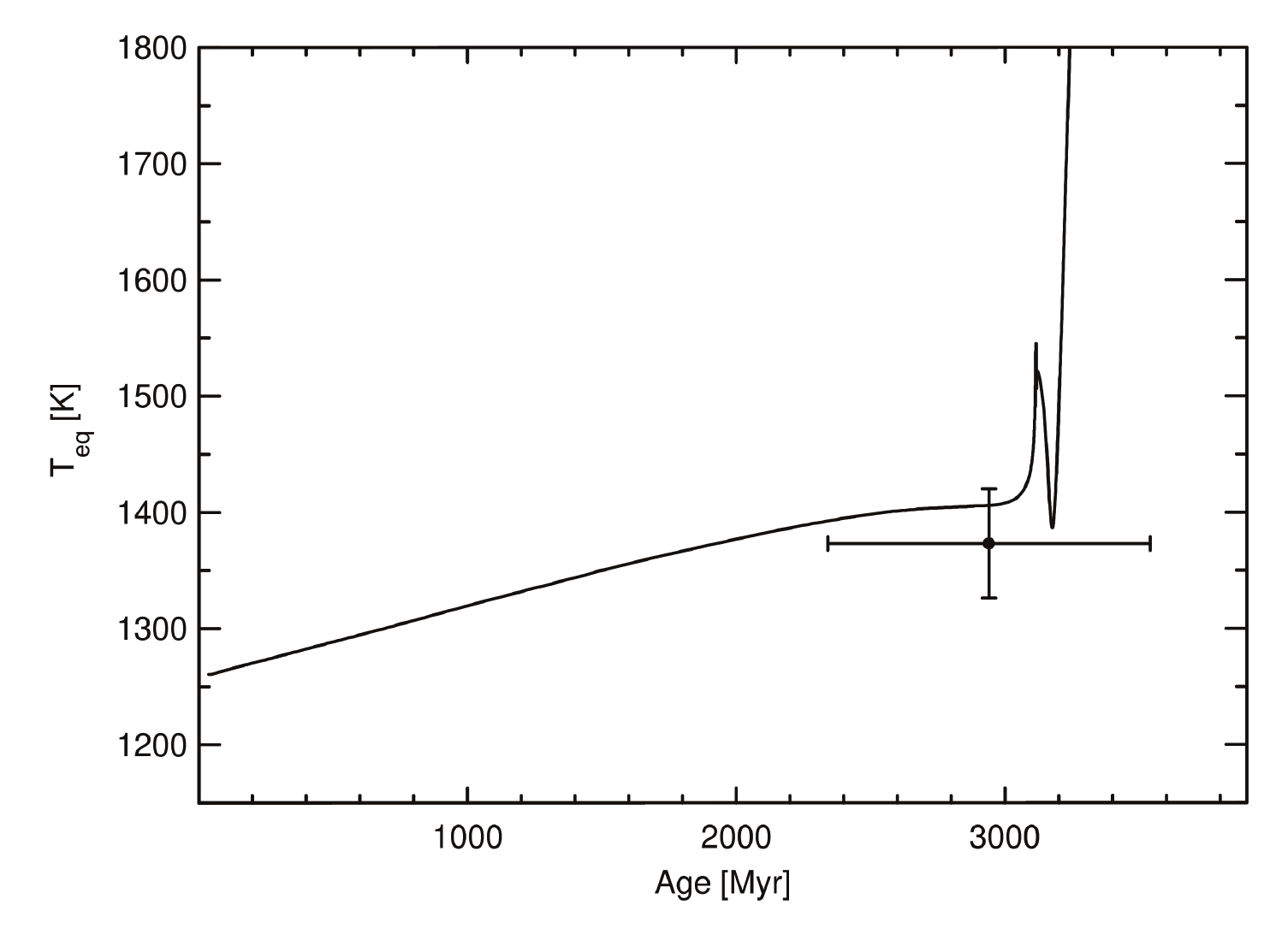}
    \caption{Equilibrium temperature as a function of the system age (solid line), computed using the closest model in the PARSEC grid (M$_\star$ = 1.45 \msun; Z$_\star$ = 0.20; see Figure \ref{fig:HR_diagram}). The position of TOI-375 b is overplotted (black dot).}
    \label{fig:Teq_age}
\end{figure}

\section{Summary} \label{sec:con}

In this work, we analyzed the radial velocity and transit photometry data of the post main sequence star TOI-375. Our best-fit model includes three Keplerian components corresponding to a hot, a warm and a cold Jupiter.

TOI-375 b is a hot Jupiter ($T_{eq} \approx 1373$) in a close-in orbit ($a \approx 0.1$\,au, $P \approx 9.45$ days) with low orbital eccentricity ($e \approx 0.09$) and well-constrained radius and mass from photometry and radial velocity measurements ($R_p \approx 0.96$ and $M_p \approx 0.74$). The two additional companions, TOI-375 c and TOI-375 d, orbit at larger separations ($a \approx 0.52$ and $0.98$\,au) with minimum masses of approximately 2.04 and 1.35~$M_{\mathrm{J}}$, respectively. No transits have been detected with a high degree of confidence for these outer planets, and their radii, orbital inclinations are unconstrained.

The system architecture is consistent with a formation scenario involving disk-driven migration. Additional data are required to rule out high-eccentricity migration pathways. Furthermore, our interior modeling of TOI-375 b supports formation via core accretion, in line with current planet formation theory.

TOI-375 is the first confirmed system with three or more fully evolved Jupiter-mass planets where at least one of the planets is transiting, making it a valuable case study for future work.

\begin{acknowledgements}
Y.R.\ would like to thank the Millennium Institute of Astrophysics (MAS) and FONDECYT for supporting this research.
N.E.\ would like to thank the Gruber Foundation for its generous support to this research.
R.B. acknowledges support from FONDECYT Project 1241963 and from ANID -- Millennium  Science  Initiative -- ICN12\_009.
A.J.\ acknowledges support from FONDECYT project 1171208 and by the Ministry for the Economy, Development, and Tourism's Programa Iniciativa Cient\'{i}fica Milenio through grant IC\,120009, awarded to the Millennium Institute of Astrophysics (MAS). 
M.B.\ acknowledges CONICYT-GEMINI grant 32180014.
T.D.\ acknowledges support from MIT’s Kavli Institute as a Kavli postdoctoral fellow.
Resources supporting this work were provided by the NASA High-End Computing (HEC) Program through the NASA Advanced Supercomputing (NAS) Division at Ames Research Center for the production of the SPOC data products.
%
This work makes use of observations from the \textit{LCOGT} network. Part of the LCOGT telescope time was granted by NOIRLab through the Mid-Scale Innovations Program (MSIP). MSIP is funded by NSF.
This research has made use of the Exoplanet Follow-up Observation Program (ExoFOP; DOI: 10.26134/ExoFOP5) website, which is operated by the California Institute of Technology, under contract with the National Aeronautics and Space Administration under the Exoplanet Exploration Program.
Funding for the \textit{TESS} mission is provided by NASA's Science Mission Directorate. KAC acknowledges support from the TESS mission via subaward s3449 from MIT.
This work has made use of data from the European Space Agency (ESA) mission Gaia (\url{https:
//www.cosmos.esa.int/gaia}), processed by the Gaia Data
Processing and Analysis Consortium (DPAC, \url{ https://www.cosmos.esa.int/web/gaia/dpac/consortium}). Funding for the DPAC has been provided by national ins
titutions, in particular
the institutions participating in the Gaia Multilateral
Agreement. 
This research made use of \textsf{exoplanet} \citep{exoplanet:exoplanet} and its
dependencies \citep{exoplanet:astropy13, exoplanet:astropy18,
exoplanet:espinoza18, exoplanet:exoplanet, exoplanet:kipping13,
exoplanet:luger18, exoplanet:pymc3, exoplanet:theano}.
We acknowledge the use of the following facilities: \textit{TESS}, FEROS/MPG 2.2m, NRES/LCOGT 1m, CTIO 1.5m/CHIRON, SINISTRO/LCOGT 1m, and MEarth-South.  

We also acknowledge the use of the following software: CERES \citep{brahm:2017:ceres,jordan:2014}, ZASPE \citep{brahm:2016:zaspe,brahm:2015}, SPECIES \citep{species}, \textsf{radvel} \citep{fulton:2018}, \textsf{batman} \citep{kreidberg:2015}, \textsf{MultiNest} \citep{multinest}, \textsf{exoplanet} \citep{exoplanet:exoplanet}, \textsf{juliet} \citep{espinoza:juliet}, AstroImageJ \citep{Collins:2017}, TAPIR \citep{Jensen:2013}, and \textsf{exostriker} \citep{2019ascl.soft06004T}.


\end{acknowledgements}

%
\bibliographystyle{bibtex/aa} 
\bibliography{toi375_refs} 

\clearpage
\newpage
\begin{appendix}
\section{Online Tables.}

\setcounter{table}{0}
\begin{table*}[ht]
    \centering
    \begin{tabular}{crrcccrrc}
        \hline
BJD-2458000 & RV(m/s) & $RV_{error}$ & Instrument &                       & BJD-2458000 & RV(m/s) & $RV_{error}$ & Instrument \\ \hline
521.54746   & 25825.8 & 6.9          & FEROS      & \vline & 644.92543   & 25843.2 & 9.6          & FEROS      \\
527.56495   & 25925.1 & 8.6          & FEROS      & \vline & 652.91304   & 25755.9 & 14.6         & FEROS      \\
542.52453   & 25815.3 & 7.7          & FEROS      & \vline & 654.88749   & 25831.9 & 6.7          & FEROS      \\
543.52255   & 25851.0 & 5.6          & FEROS      & \vline & 657.88536   & 25903.2 & 6.1          & FEROS      \\
544.55314   & 25886.0 & 6.3          & FEROS      & \vline & 663.91369   & 25885.4 & 8.8          & FEROS      \\
545.53503   & 25880.1 & 6.9          & FEROS      & \vline & 669.87593   & 25867.9 & 6.7          & FEROS      \\
546.52085   & 25860.2 & 7.1          & FEROS      & \vline & 676.89385   & 25969.9 & 8.6          & FEROS      \\
547.51570   & 25812.4 & 6.2          & FEROS      & \vline & 677.87102   & 25948.6 & 7.5          & FEROS      \\
548.51955   & 25784.2 & 5.9          & FEROS      & \vline & 705.93076   & -2326.8 & 10.8         & CHIRON*    \\
549.52650   & 25768.4 & 6.1          & FEROS      & \vline & 711.91847   & -2374.9 & 19.1         & CHIRON*    \\
550.51710   & 25776.8 & 6.2          & FEROS      & \vline & 712.92511   & -2346.9 & 11.2         & CHIRON*    \\
551.51510   & 25805.3 & 7.4          & FEROS      & \vline & 718.81491   & 25978.9 & 6.0          & FEROS      \\
553.53751   & 25854.3 & 9.7          & FEROS      & \vline & 721.83211   & 26008.5 & 6.5          & FEROS      \\
554.51806   & 25881.4 & 9.7          & FEROS      & \vline & 721.84588   & -2331.7 & 11.6         & CHIRON*    \\
555.54327   & 25862.3 & 6.2          & FEROS      & \vline & 722.79761   & 26047.3 & 5.9          & FEROS      \\
556.54585   & 25827.8 & 6.4          & FEROS      & \vline & 723.78940   & -2308.6 & 15.3         & CHIRON*    \\
558.51625   & 25759.2 & 7.8          & FEROS      & \vline & 723.83252   & 26091.9 & 7.1          & FEROS      \\
559.54811   & 25774.8 & 7.1          & FEROS      & \vline & 724.84107   & 26058.2 & 7.6          & FEROS      \\
560.51685   & 25812.0 & 6.5          & FEROS      & \vline & 725.72397   & 26055.7 & 6.6          & FEROS      \\
566.54816   & 25812.8 & 5.8          & FEROS      & \vline & 728.87905   & -2401.1 & 12.2         & CHIRON*    \\
568.54820   & 25812.7 & 6.2          & FEROS      & \vline & 734.89998   & -2300.9 & 12.5         & CHIRON*    \\
572.54079   & 25936.5 & 9.0          & FEROS      & \vline & 784.77100   & 25820.8 & 6.7          & FEROS      \\
574.50180   & 25878.6 & 8.1          & FEROS      & \vline & 794.68403   & 25822.1 & 6.0          & FEROS      \\
578.49890   & 8024.0  & 21.1         & CHIRON     & \vline & 798.70163   & 25900.0 & 6.5          & FEROS      \\
580.48711   & 8115.9  & 15.8         & CHIRON     & \vline & 798.73782   & 25888.9 & 7.4          & FEROS      \\
582.48064   & 8144.0  & 12.1         & CHIRON     & \vline & 799.68013   & 25935.0 & 6.5          & FEROS      \\
637.92845   & 25955.7 & 7.1          & FEROS      & \vline & 811.63493   & 8131.8  & 16.6         & CHIRON     \\
638.91690   & 25971.7 & 7.5          & FEROS      & \vline & 825.65294   & 8162.3  & 13.9         & CHIRON     \\
639.93184   & 25951.0 & 8.6          & FEROS      & \vline & 899.53159   & 7912.8  & 24.4         & CHIRON     \\
640.93119   & 25921.2 & 6.9          & FEROS      & \vline & 905.52999   & 8070.2  & 16.3         & CHIRON     \\
641.92771   & 25900.8 & 8.1          & FEROS      & \vline & 912.50716   & 8106.0  & 15.1         & CHIRON     \\
642.93776   & 25829.3 & 6.2          & FEROS      & \vline & 920.50064   & 8116.0  & 17.7         & CHIRON     \\
643.93254   & 25843.3 & 7.7          & FEROS      & \vline &             &         &              &            \\ \hline
    \end{tabular}
    \caption{Radial velocities for TOI-375. Data taken with CHIRON between August and September of 2019 was marked with an asterisk (*)  }
    \label{tab:rvs}
\end{table*}

\clearpage

\begin{table*}[ht!]
\centering
\begin{tabular}{cccccccc}
\hline
\hline
\label{tab:rv_models}

model      & $^{TOI-375c}_{orbit}$ & $^{TOI-375d}_{orbit}$ & RV\par{}$^{linear}_{trend}$          & log(z)    &  $N_{params}$    &AIC  & BIC   \\
\hline
\textbf{A}         & \textbf{circular}            & \textbf{circular}      & \textbf{no}    & \textbf{-321.3}   &  \textbf{18} & \textbf{678.6}&\textbf{718.0}  \\
B & circular  & circular   & yes &                -328.4    &  20& 696.8&740.6\\
C          & circular& elliptic& no       & -322.4      &  20  &684.8 &728.6 \\
D          & circular  & elliptic         & yes         & -328.0   &22 &700.0 &748.2      \\
E          & elliptic     & circular         & no        & -324.4       &20&688.8 &732.6\\
F          & elliptic     & circular      & yes          &    -332.4     &22& 708.8&757.0   \\
G          & elliptic     & elliptic           & no     & -323.4         & 22& 690.8&739.0\\
H          & elliptic     & elliptic           & yes         & -326.5     &24& 701.0&753.6\\  
\hline
model      & $^{TOI-375c}_{orbit}$ &$^{TOI-375d}_{orbit}$ & $^{TOI-375e}_{orbit}$      & log(z)   &     $N_{params}$ & AIC &BIC     \\
\hline
I&circular&circular&circular&-323.6&21&689.2&735.2\\
J&circular&circular&elliptic&-326.2&23&698.4&748.8\\
K&circular&elliptic&circular&-322.9&23&691.8&742.2\\
L&circular&elliptic&elliptic&-331.5&25&713.0&767.7\\
M&elliptic&circular&circular&-322.6&23&691.2&741.6\\
N&elliptic&circular&elliptic&-325.9&25&701.8&756.5\\
O&elliptic&elliptic&circular&-324.9&25&699.8&754.5\\
P&elliptic&elliptic&elliptic&-332.5&27&719.0&778.1\\

\end{tabular}
\caption{Comparison table between different radial velocity models. Models A trough H consists of 3 planets with uniform priors for the period of the 2 outer planets:100-200 and 200-500 days respectively, and with or without a linear trend. Model I trough P consist of 4 planets with uniform priors for the period of the outer 3 planets:10-100, 100-200 and 200-500 days respectively. All models have a gaussian prior for the  epoch and period of TOI-375b  provided by the light curve model. }
\end{table*}

\tiny
\begin{table*}[ht!]
    \centering
    {\small
    \begin{tabular}{l c r}
        \hline
        \hline
Parameter & Prior &  Posterior \\
\hline
$\rho_*$ & $\mathcal{N}(83.24, 12.0)$ & $75.20^{+8.35}_{-8.75}$ \\
$P_b$ & $\mathcal{U}(9.4,9.5)$ & $9.454693^{+0.000017}_{-0.000018}$ \\
$t0_b$ & $\mathcal{U}(2458329.2,2458329.4)$ & $2458329.2911^{+0.0022}_{-0.0022}$ \\
$p_b$ & $\mathcal{U}(0.0,1.0)$ & $0.03401^{+0.00044}_{-0.00042}$ \\
$b_b$ & $\mathcal{U}(0.0,1.0)$ & $0.177^{+0.138}_{-0.114}$ \\
$K_b$ & $\mathcal{U}(10.0,100.0)$ & $56.31^{+3.37}_{-3.42}$ \\
$\sqrt{e}\cos\omega_b$ & $\mathcal{U}(-0.5,0.5)$ & $0.267^{+0.061}_{-0.085}$ \\
$\sqrt{e}\sin\omega_b$ & $\mathcal{U}(-0.5,0.5)$ & $0.037^{+0.126}_{-0.126}$ \\
$P_c$ & $\mathcal{U}(100.0,200.0)$ & $115.54^{+1.96}_{-1.67}$ \\
$t0_c$ & $\mathcal{U}(2458429.0,2458529.0)$ & $2458520.5353^{+2.13}_{-2.21}$ \\
$K_c$ & $\mathcal{U}(10.0,100.0)$ & $69.08^{+6.69}_{-6.57}$ \\
$P_d$ & $\mathcal{U}(200.0,450.0)$ & $297.13^{+28.60}_{-17.83}$ \\
$t0_d$ & $\mathcal{U}(2458729.0,2458929.0)$ & $2458790.7853^{+10.67}_{-10.09}$ \\
$K_d$ & $\mathcal{U}(10.0,100.0)$ & $33.38^{+7.14}_{-6.60}$ \\
$q_1^{MEarth}$ & $\mathcal{U}(0.0,1.0)$ & $0.553^{+0.293}_{-0.310}$ \\
$q_2^{MEarth}$ & $\mathcal{U}(0.0,1.0)$ & $0.415^{+0.346}_{-0.288}$ \\
$mflux^{MEarth}$ & $\mathcal{U}(-0.1,0.1)$ & $-0.00424^{+0.00155}_{-0.00181}$ \\
$\sigma_w^{MEarth}$ & $\mathcal{U}(0.0,2000.0)$ & $874.78^{+66.33}_{-67.80}$ \\
$\theta_0^{MEarth}$ & $\mathcal{U}(-0.1,0.1)$ & $-0.00113^{+0.00075}_{-0.00097}$ \\
$\theta_1^{MEarth}$ & $\mathcal{U}(-0.1,0.1)$ & $-0.00020^{+0.00004}_{-0.00004}$ \\
$GP_\sigma^{MEarth}$ & $\mathcal{J}(1\mathrm{e}{-}6,1.0)$ & $0.00033^{+0.00042}_{-0.00012}$ \\
$GP{\rho}^{MEarth}$ & $\mathcal{J}(0.001,1.0)$ & $0.0203^{+0.0579}_{-0.0111}$ \\
$q_1^{LCOGT}$ & $\mathcal{U}(0.0,1.0)$ & $0.438^{+0.343}_{-0.297}$ \\
$q_2^{LCOGT}$ & $\mathcal{U}(0.0,1.0)$ & $0.477^{+0.340}_{-0.321}$ \\
$mflux^{LCOGT}$ & $\mathcal{U}(-0.1,0.1)$ & $-0.0386^{+0.0257}_{-0.0217}$ \\
$\sigma_w^{LCOGT}$ & $\mathcal{U}(0.0,2000.0)$ & $767.06^{+144.65}_{-149.03}$ \\
$\theta_0^{LCOGT}$ & $\mathcal{U}(-0.1,0.1)$ & $-0.0397^{+0.0271}_{-0.0239}$ \\
$\theta_1^{LCOGT}$ & $\mathcal{U}(-0.1,0.1)$ & $-0.00691^{+0.00472}_{-0.00473}$ \\
$q_1^{TESS}$ & $\mathcal{U}(0.0,1.0)$ & $0.620^{+0.221}_{-0.219}$ \\
$q_2^{TESS}$ & $\mathcal{U}(0.0,1.0)$ & $0.153^{+0.155}_{-0.095}$ \\
$mflux^{TESS-S1}$ & $\mathcal{U}(-0.01,0.01)$ & $-0.000013^{+0.000041}_{-0.000040}$ \\
$\sigma_w^{TESS-S1}$ & $\mathcal{U}(0.0,1000.0)$ & $42.42^{+34.14}_{-28.34}$ \\
$GP\sigma^{TESS-S1}$ & $\mathcal{J}(1\mathrm{e}{-}6,1.0)$ & $0.000168^{+0.000032}_{-0.000025}$ \\
$GP{\rho}^{TESS-S1}$ & $\mathcal{J}(0.001,1.0)$ & $0.415^{+0.130}_{-0.099}$ \\
$mflux^{TESS-S2}$ & $\mathcal{U}(-0.01,0.01)$ & $-0.000107^{+0.000080}_{-0.000076}$ \\
$\sigma_w^{TESS-S2}$ & $\mathcal{U}(0.0,1000.0)$ & $30.84^{+30.30}_{-21.31}$ \\
$GP\sigma^{TESS-S2}$ & $\mathcal{J}(1\mathrm{e}{-}6,1.0)$ & $0.000207^{+0.000056}_{-0.000042}$ \\
$GP{\rho}^{TESS-S2}$ & $\mathcal{J}(0.001,1.0)$ & $0.615^{+0.211}_{-0.173}$ \\
$mflux^{TESS-S13}$ & $\mathcal{U}(-0.01,0.01)$ & $-0.000128^{+0.000016}_{-0.000016}$ \\
$\sigma_w^{TESS-S13}$ & $\mathcal{U}(0.0,1000.0)$ & $57.73^{+53.19}_{-39.61}$ \\
$mflux^{TESS-S27}$ & $\mathcal{U}(-0.01,0.01)$ & $-0.000109^{+0.000019}_{-0.000020}$ \\
$\sigma_w^{TESS-S27}$ & $\mathcal{U}(0.0,1000.0)$ & $116.67^{+84.11}_{-76.70}$ \\
$mflux^{TESS-S28}$ & $\mathcal{U}(-0.01,0.01)$ & $-0.000209^{+0.000022}_{-0.000022}$ \\
$\sigma_w^{TESS-S28}$ & $\mathcal{U}(0.0,1000.0)$ & $82.65^{+73.35}_{-56.90}$ \\
$mflux^{TESS-S39}$ & $\mathcal{U}(-0.01,0.01)$ & $-0.000136^{+0.000017}_{-0.000017}$ \\
$\sigma_w^{TESS-S39}$ & $\mathcal{U}(0.0,1000.0)$ & $61.65^{+56.54}_{-42.10}$ \\
$mflux^{TESS-S66}$ & $\mathcal{U}(-0.01,0.01)$ & $-0.000230^{+0.000024}_{-0.000024}$ \\
$\sigma_w^{TESS-S66}$ & $\mathcal{U}(0.0,1000.0)$ & $75.10^{+67.83}_{-51.53}$ \\
$mflux^{TESS-S67}$ & $\mathcal{U}(-0.01,0.01)$ & $-0.000308^{+0.000027}_{-0.000027}$ \\
$\sigma_w^{TESS-S67}$ & $\mathcal{U}(0.0,1000.0)$ & $78.44^{+74.80}_{-53.92}$ \\
$mflux^{TESS-S68}$ & $\mathcal{U}(-0.01,0.01)$ & $-0.000172^{+0.000021}_{-0.000020}$ \\
$\sigma_w^{TESS-S68}$ & $\mathcal{U}(0.0,1000.0)$ & $84.21^{+72.48}_{-56.95}$ \\
$\mu^{FEROS}$ & $\mathcal{U}(25800.0,26000.0)$ & $25919.17^{+3.58}_{-3.76}$ \\
$\sigma_w^{FEROS}$ & $\mathcal{U}(10.0,40.0)$ & $18.06^{+2.83}_{-2.43}$ \\
$\mu^{CHIRON-1}$ & $\mathcal{U}(7900.0,8200.0)$ & $8119.16^{+13.62}_{-13.33}$ \\
$\sigma_w^{CHIRON-1}$ & $\mathcal{U}(10.0,40.0)$ & $23.30^{+9.25}_{-8.23}$ \\
$\mu^{CHIRON-2}$ & $\mathcal{U}(-2600.0,-2200.0)$ & $-2445.73^{+9.23}_{-9.31}$ \\
$\sigma_w^{CHIRON-2}$ & $\mathcal{U}(10.0,40.0)$ & $16.48^{+7.79}_{-4.57}$ \\
\hline
\end{tabular}}
\caption{Complete set of priors used in the \textsf{juliet} runs and their corresponding posterior distributions. Parameter units follow the \textsf{juliet} documentation \citep{espinoza:juliet}. }
\label{table:full}
\end{table*}

\clearpage
\newpage

\section{Online Figures}
\setcounter{figure}{0}

\begin{figure*}[htp]
\centering
\setlength{\lineskip}{0pt}
\makebox[0pt][r]{\makebox[1em][l]{A}}%
\includegraphics[width=0.33\textwidth,valign=t]{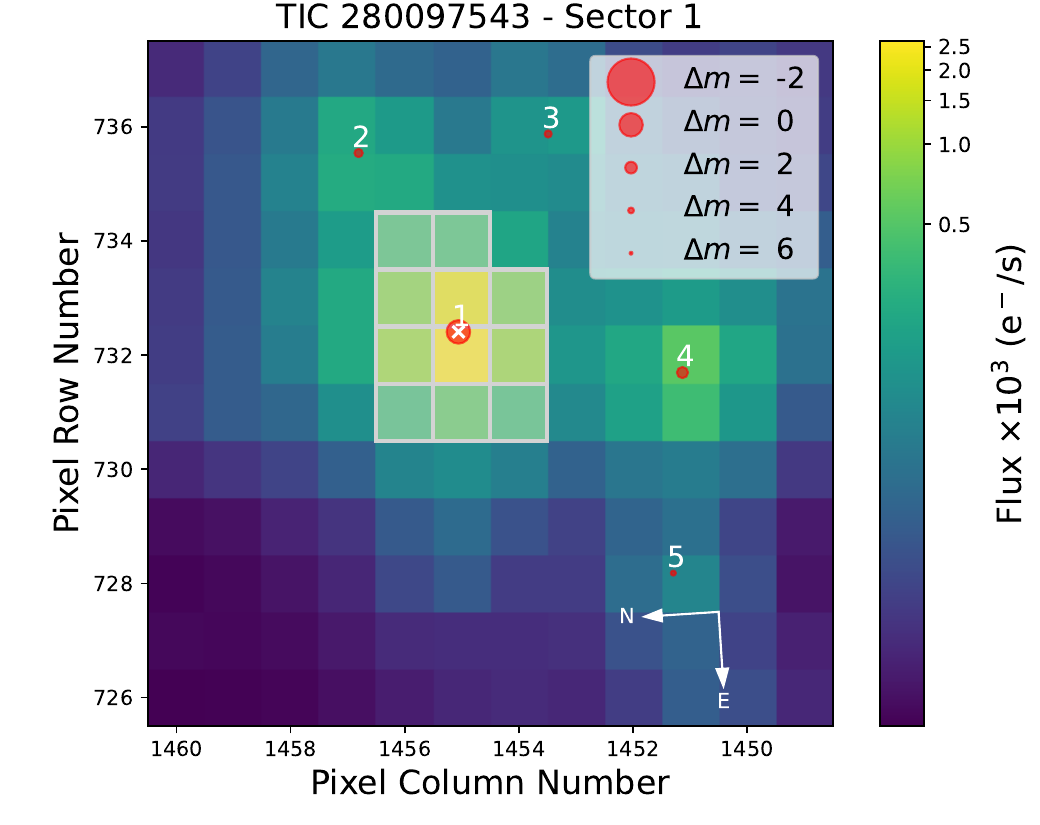}%
\includegraphics[width=0.33\textwidth,valign=t]{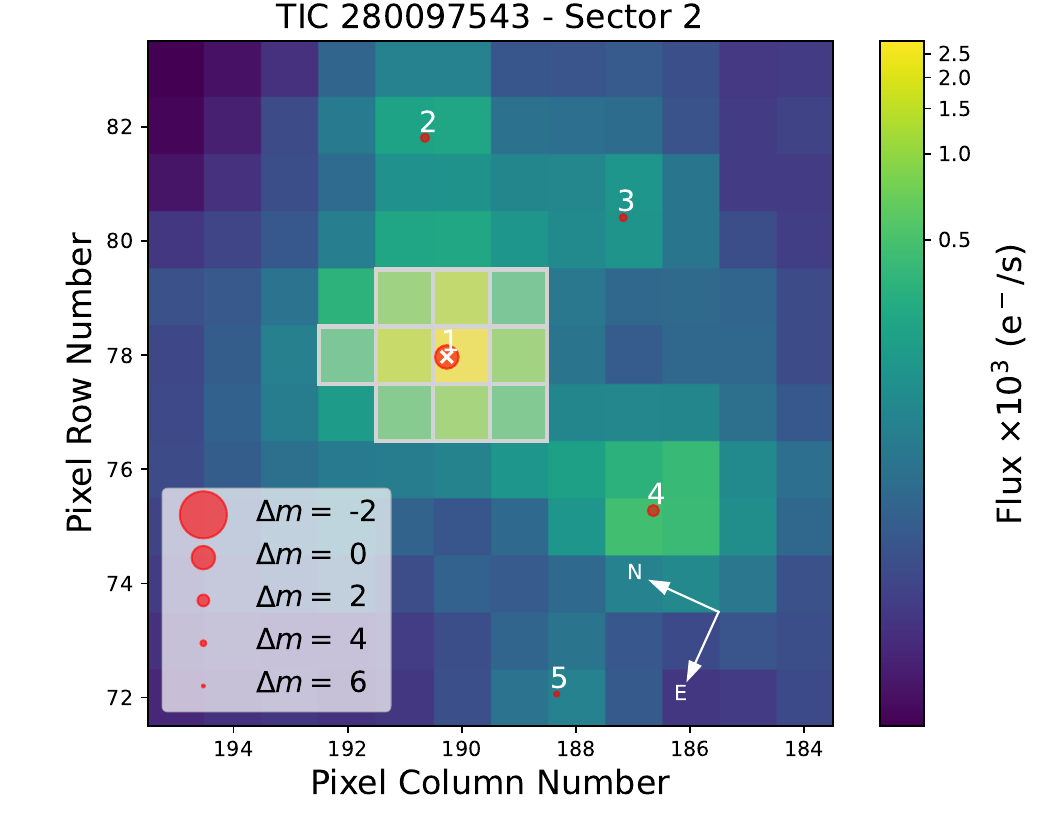}%
\includegraphics[width=0.33\textwidth,valign=t]{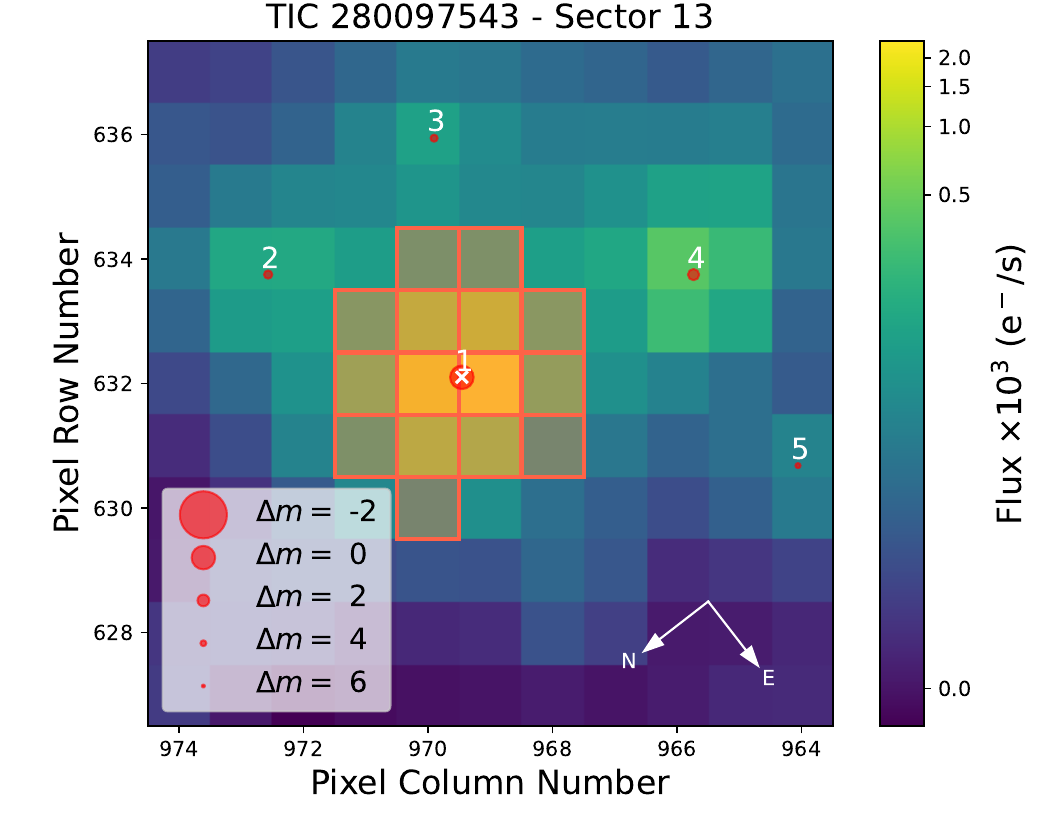}\\
\makebox[0pt][r]{\makebox[1em][l]{B}}%
\includegraphics[width=0.33\textwidth,valign=t]{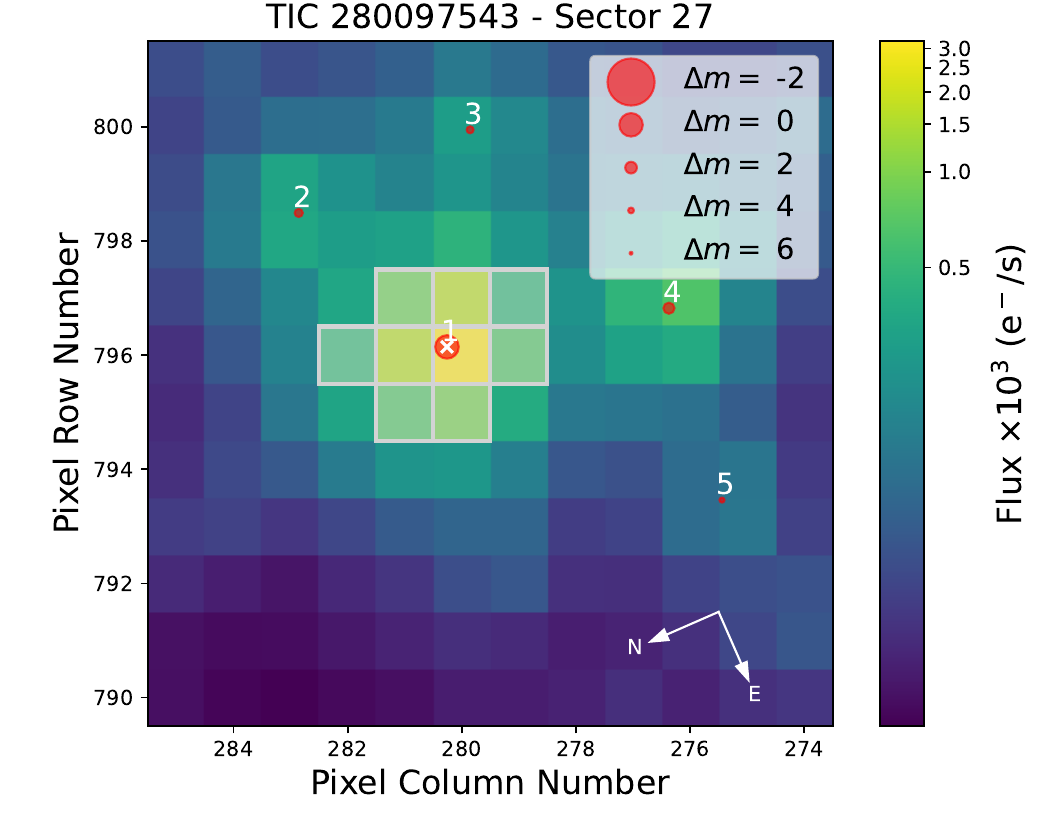}%
\includegraphics[width=0.33\textwidth,valign=t]{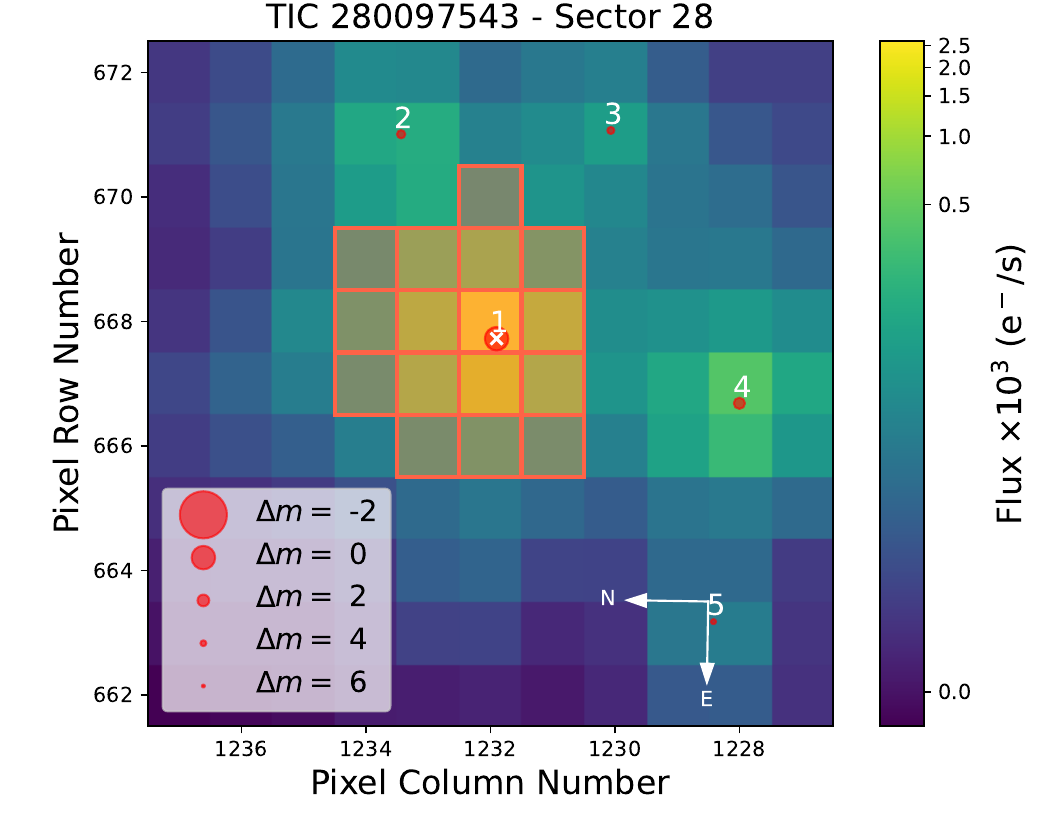}%
\includegraphics[width=0.33\textwidth,valign=t]{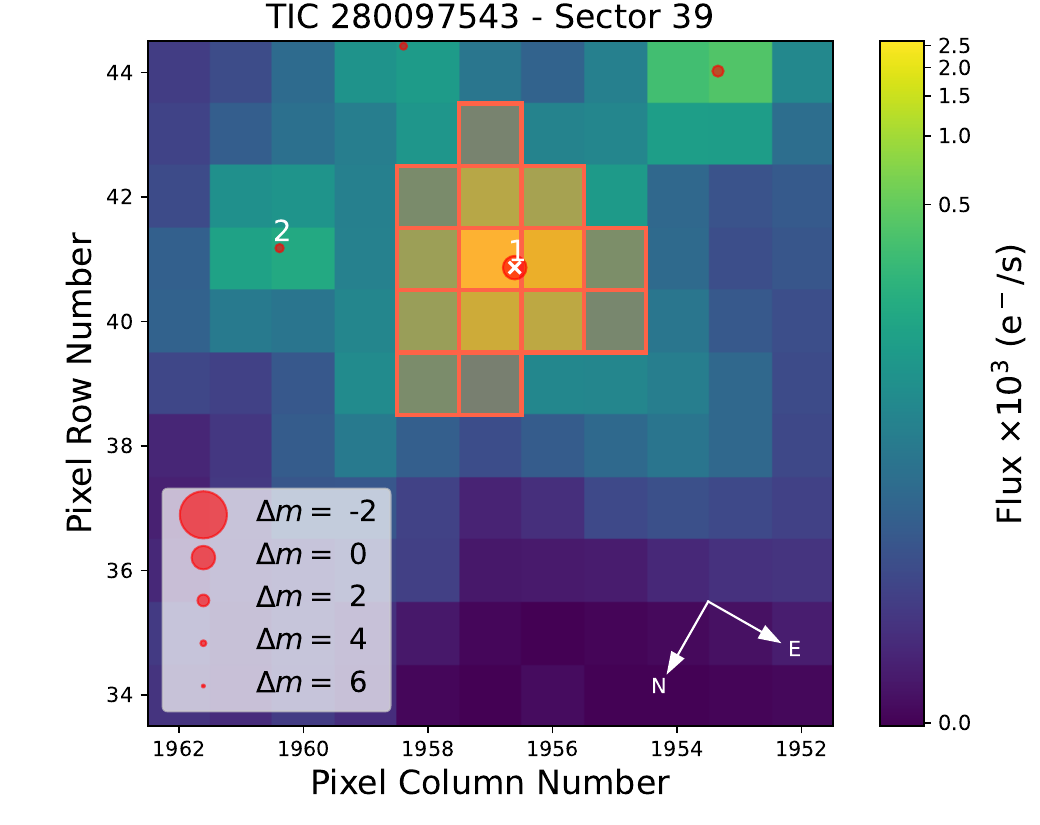}\\
\makebox[0pt][r]{\makebox[1em][l]{C}}%
\includegraphics[width=0.33\textwidth,valign=t]{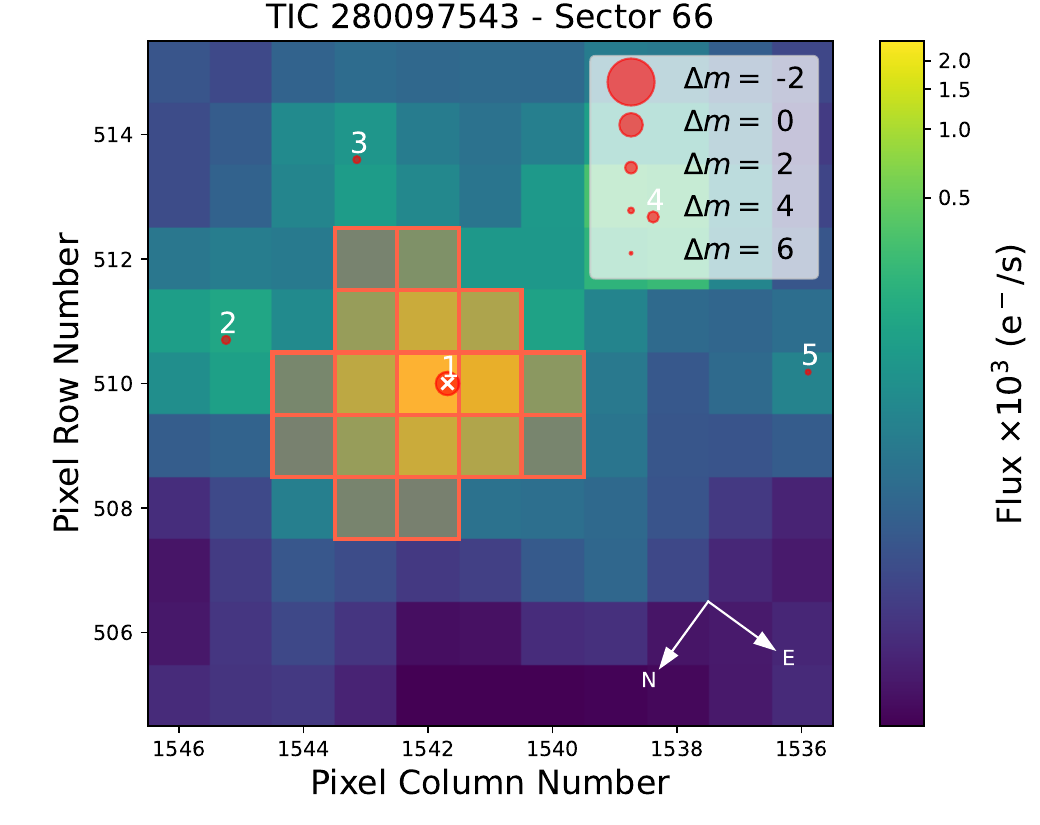}%
\includegraphics[width=0.33\textwidth,valign=t]{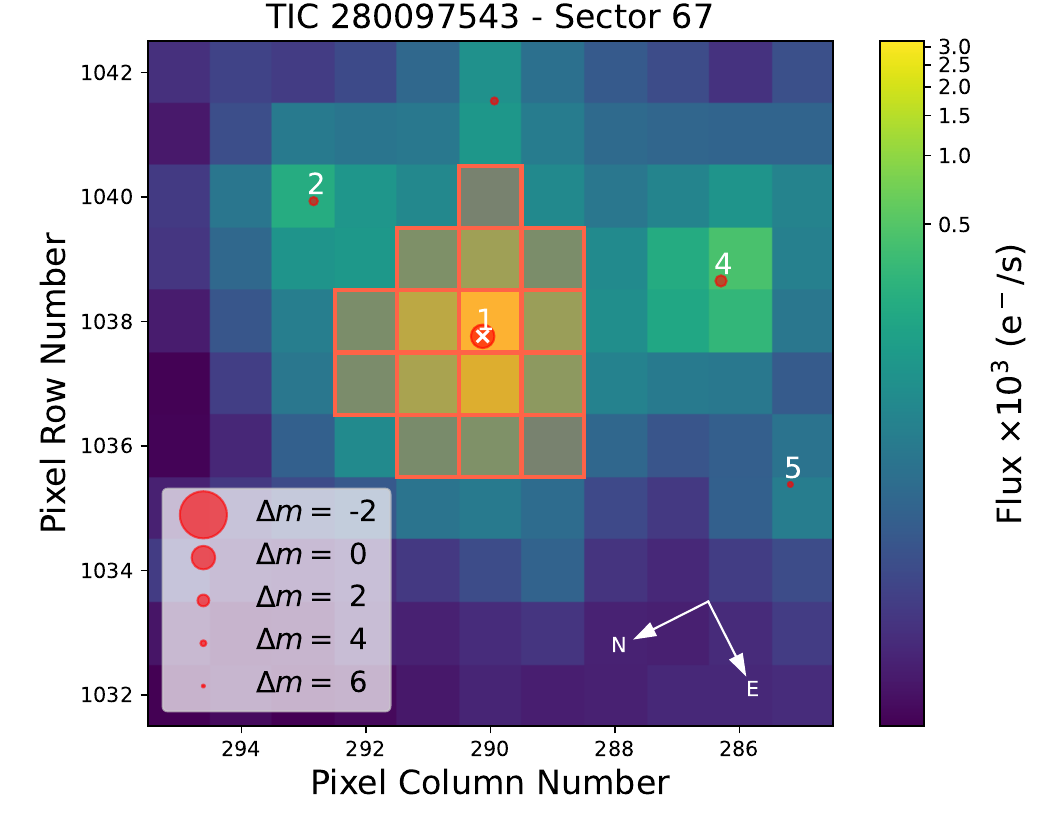}%
\includegraphics[width=0.33\textwidth,valign=t]{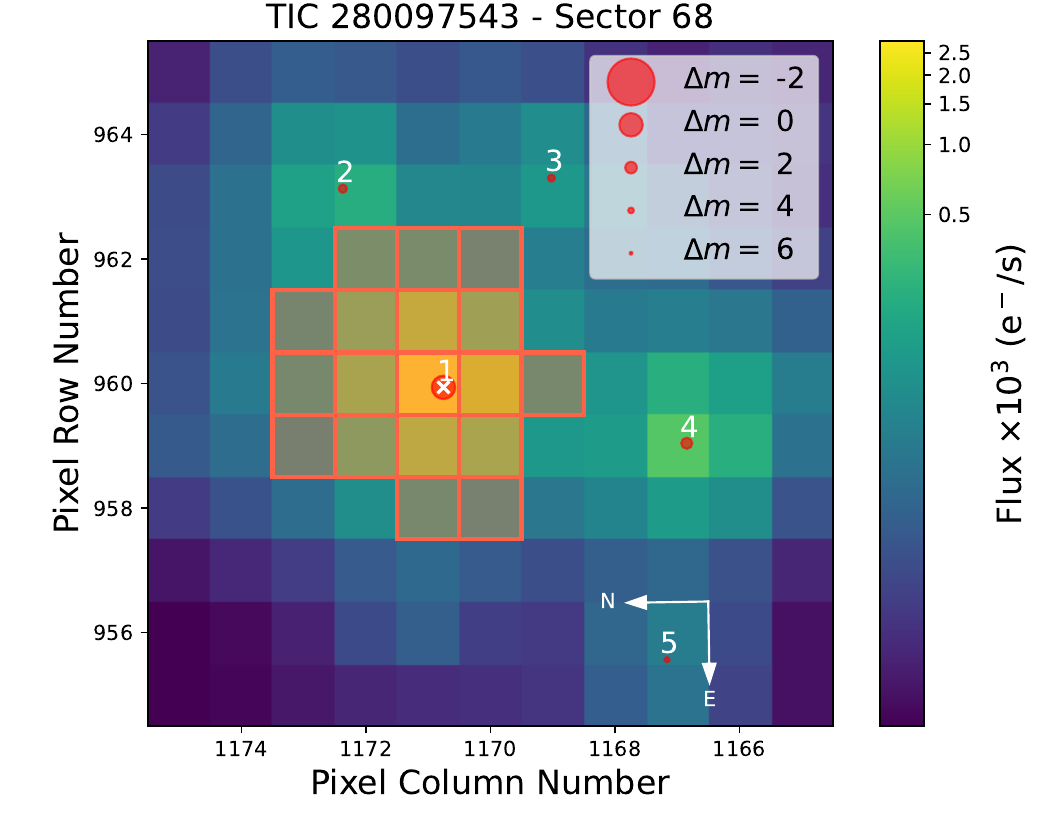}

\caption{Target pixel file (TPF) for TOI-375 sectors; the target star is marked with a white cross on top of a red point, and are numbered as 1. Smaller numbered red points are the closest stars to the target (drawn from Gaia) with Gaia magnitude differences with the target of $|\Delta G| < 6$. There are no nearby sources of contamination for TOI-375. Plot made using \textsf{tpfplotter}.\label{fig:tpf}}

\end{figure*}

\FloatBarrier 
\clearpage

\end{appendix}
\end{document}